\documentclass[11pt,onecolumn,draftclsnofoot]{IEEEtran}
\usepackage{graphicx,accents}
\usepackage[cmex10]{amsmath}
\usepackage{amsfonts,amssymb,latexsym,cite,ifthen,url}
\usepackage[usenames,dvipsnames]{color}
\usepackage[hang, font=footnotesize]{caption}
\newboolean{SEP_FIG_CAPS}\setboolean{SEP_FIG_CAPS}{false}
\newboolean{ONE_COLUMN}\setboolean{ONE_COLUMN}{true}
\ifthenelse{\boolean{SEP_FIG_CAPS}}
{
 
 \newcommand{\putTable}[3]{\begin{table}[p]
  			    \centering
		            #3
            		    \caption{}
     			    \label{tab:#1}
			  \end{table}
			  \clearpage}
 \newcommand{\capFrag}[2]{\noindent Fig.~\ref{fig:#1}. #2 \medskip\\}
 \newcommand{\capTable}[2]{\noindent Tab.~\ref{tab:#1}. #2 \medskip\\}
}
{
 
 \newcommand{\putTable}[3]{\begin{table}[h!]
  			    \centering
		            #3
     			    \caption{#2}
     			    \label{tab:#1}
			  \end{table} }
 \newcommand{\capFrag}[2]{}
 \newcommand{\capTable}[2]{}
}

 \newcommand{\ovec}[1]{\ensuremath{\Bar{\boldsymbol{#1}}}}
 \newcommand{\hvec}[1]{\ensuremath{\Hat{\boldsymbol{#1}}}}
 
 \renewcommand{\vec}[1]{\ensuremath{\boldsymbol{#1}}}

 \DeclareMathOperator{\vect}{vec}

 \DeclareMathOperator*{\argmax}{argmax}

 \renewcommand{\eqref}[1]{(\ref{eq:#1})}
 \newcommand{\Eqref}[1]{Equation~(\ref{eq:#1})}
 \newcommand{\Figref}[1]{Figure~\ref{fig:#1}}
 \newcommand{\figref}[1]{Fig.~\ref{fig:#1}}
 \newcommand{\tabref}[1]{Table~\ref{tab:#1}}
 \newcommand{\secref}[1]{Section~\ref{sec:#1}}
 
 \newcommand{\appref}[1]{Appendix~\ref{app:#1}}

 \newcounter{comment}[section]
 
 \newcounter{texthead}[section]

 \newcommand{\fwd}[1]{\accentset{\rightharpoonup}{#1}}
 \newcommand{\bwd}[1]{\accentset{\leftharpoonup}{#1}}
 
 \newcommand{\eps}{\varepsilon}

 \newcommand{\mf}[1]{\mathfrak{#1}}
 \renewcommand{\vect}[1]{\vec{#1}^{(t)}}
 \newcommand{\nt}[1]{{#1}_n^{(t)}}
 \newcommand{\msg}[2]{\nu_{{#1} \to {#2}}}
 
 \newcounter{threshcounter}
 \renewcommand{\thethreshcounter}{D\arabic{threshcounter}}
 \newcommand{\threshcnt}[1]{\refstepcounter{threshcounter} \label{#1} \thetag{\thethreshcounter}}
 \newcounter{algcounter}
 \renewcommand{\thealgcounter}{A\arabic{algcounter}}
 \newcommand{\algcnt}[1]{\refstepcounter{algcounter} \label{#1} \thetag{\thealgcounter}} \newcounter{emcounter}
 \renewcommand{\theemcounter}{E\arabic{emcounter}}
 \newcommand{\emcnt}[1]{\refstepcounter{emcounter} \label{#1} \thetag{\theemcounter}}

 \graphicspath{{./}{Figures/}}

\begin{document}
\setlength{\arraycolsep}{0.8mm}

%
\title{Efficient High-Dimensional Inference in the Multiple Measurement Vector Problem}
\author{Justin Ziniel~\IEEEmembership{Student Member,~IEEE} and Philip Schniter~\IEEEmembership{Senior Member,~IEEE}
\thanks{The authors are with the Department of Electrical and Computer Engineering, The Ohio State University, Columbus, Ohio.  E-mail: \{zinielj, schniter\}@ece.osu.edu.  Please direct all correspondence to Justin Ziniel.}
\thanks{Work supported in part by NSF grant CCF-1018368, DARPA/ONR grant N66001-10-1-4090, and an allocation of computing time from the Ohio Supercomputer Center.}
}

\maketitle

\begin{abstract}
In this work, a Bayesian approximate message passing algorithm is proposed for solving the multiple measurement vector (MMV) problem in compressive sensing, in which a collection of sparse signal vectors that share a common support are recovered from undersampled noisy measurements.  The algorithm, AMP-MMV, is capable of exploiting temporal correlations in the amplitudes of non-zero coefficients, and provides soft estimates of the signal vectors as well as the underlying support.  Central to the proposed approach is an extension of recently developed approximate message passing techniques to the amplitude-correlated MMV setting.  Aided by these techniques, AMP-MMV offers a computational complexity that is linear in all problem dimensions.  In order to allow for automatic parameter tuning, an expectation-maximization algorithm that complements AMP-MMV is described.  Finally, a detailed numerical study demonstrates the power of the proposed approach and its particular suitability for application to high-
dimensional problems.
\end{abstract}

\section{Introduction}
\label{sec:introduction}
As the field of compressive sensing (CS) \cite{CRT2006, D2006, CW2008} matures, researchers have begun to explore numerous extensions of the classical sparse signal recovery problem, in which a signal with few non-zero coefficients is reconstructed from a handful of incoherent linear measurements.  One such extension, known as the \emph{multiple measurement vector} (MMV) problem, generalizes the sparse signal recovery, or \emph{single measurement vector} (SMV), problem to the case where a group of measurement vectors has been obtained from a group of signal vectors that are assumed to be jointly sparse, sharing a common support.  Such a problem has many applications, including magnetoencephalography \cite{RK1998,CREK2005}, direction-of-arrival estimation \cite{TMT2010} and parallel magnetic resonance imaging (pMRI) \cite{LYL2009}.

Mathematically, given $T$ length-$M$ measurement vectors, the traditional MMV objective is to recover a collection of length-$N$ sparse vectors $\{\vect{x}\}_{t=1}^T$, when $M < N$.  Each measurement vector, $\vect{y}$, is obtained as 
\begin{equation}
  \vect{y} = \vec{A} \vect{x} + \vect{e}, \quad t = 1, \ldots, T,
  \label{eq:linear_model}
\end{equation}
where $\vec{A}$ is a known measurement matrix and $\vect{e}$ is corrupting additive noise.  The unique feature of the MMV problem is the assumption of joint sparsity: the support of each sparse signal vector $\vect{x}$ is identical.  Oftentimes, the collection of measurement vectors form a time-series, thus we adopt a temporal viewpoint of the MMV problem, without loss of generality.

A straightforward approach to solving the MMV problem is to break it apart into independent SMV problems and apply one of the many SMV algorithms.  While simple, this approach ignores valuable temporal structure in the signal that can be exploited to provide improved recovery performance.  Indeed, under mild conditions, the probability of recovery failure can be made to decay exponentially as the number of timesteps $T$ grows, when taking into account the joint sparsity \cite{ER2010}.

{Another approach (e.g., \cite{EM2009}) to the joint-sparse MMV problem is to restate \eqref{linear_model} as the block-sparse SMV model
\begin{equation}
	\ovec{y} = \vec{\mathcal{D}}\!(\!\vec{A}\!) \ovec{x} + \ovec{e},
	\label{eq:smv_linear_model}
\end{equation}
where $\ovec{y} \triangleq \big[\vec{y}^{(1)^{\textsf{T}}},\ldots,\vec{y}^{(T)^{\textsf{T}}}\big]^{{}^\textsf{T}}$, $\ovec{x} \triangleq \big[\vec{x}^{(1)^{\textsf{T}}},\ldots,\vec{x}^{(T)^{\textsf{T}}}\big]^{{}^\textsf{T}}$, $\ovec{e} \triangleq \big[\vec{e}^{(1)^{\textsf{T}}},\ldots,\vec{e}^{(T)^{\textsf{T}}}\big]^{{}^\textsf{T}}$, and $\vec{\mathcal{D}}\!(\!\vec{A}\!)$ denotes a block diagonal matrix consisting of $T$ replicates of $\vec{A}$.  In this case, $\vec{x}$ is block-sparse, where the $n^{th}$ block (for $n = 1,\ldots,N$) consists of the coefficients $\{x_n,x_{n+N},\ldots,x_{n+(T-1)N}\}$.  Equivalently, one could express \eqref{linear_model} using the matrix model
\begin{equation}
	\vec{Y} = \vec{A}\vec{X} + \vec{E},
	\label{eq:matrix_linear_model}
\end{equation}
where $\vec{Y} \triangleq \big[\vec{y}^{(1)},\ldots,\vec{y}^{(T)}\big]$, $\vec{X} \triangleq \big[\vec{x}^{(1)},\ldots,\vec{x}^{(T)}\big]$, and $\vec{E} \triangleq \big[\vec{e}^{(1)},\ldots,\vec{e}^{(T)}\big]$.  Under the matrix model, joint sparsity in \eqref{linear_model} manifests as row-sparsity in $\vec{X}$.}  Algorithms developed for the matrix MMV problem are oftentimes intuitive extensions of SMV algorithms, and therefore share a similar taxonomy.  Among the different techniques that have been proposed are mixed-norm minimization methods \cite{CREK2005, TGS2006b, HM2009, ZR2011b}, greedy pursuit methods \cite{CREK2005, TGS2006a, LBJ2010}, and Bayesian methods \cite{WR2007, TMT2010, KCJBY2011, SC2011, ZR2011}.  Existing literature suggests that greedy pursuit techniques are outperformed by mixed-norm minimization approaches, which in turn are surpassed by Bayesian methods \cite{CREK2005,WR2007,ZR2011}.

In addition to work on the MMV problem, related work has been performed on a similar problem sometimes referred to as the ``\emph{dynamic CS}'' problem \cite{V2008,AGG2009,ARG2009,VL2010,ZPS2010}.  The dynamic CS problem also shares the trait of working with multiple measurement vectors, but instead of joint sparsity, considers a situation in which the support of the signal changes slowly over time.

{Given the plethora of available techniques for solving the MMV problem, it is natural to wonder what, if any, improvements can be made.  In this work, we will primarily address two deficiencies evident in the available MMV literature.  The first deficiency is the inability of many algorithms to account for amplitude correlations in the non-zero rows of $\vec{X}$.\footnote{{Notable exceptions include \cite{KCJBY2011}, \cite{ZR2011b}, and \cite{ZR2011}, which explicitly model amplitude correlations.}}  Incorporating this temporal correlation structure is crucial, not only because many real-world signals possess such structure, but because the performance of MMV algorithms is particularly sensitive to this structure \cite{WR2007,BF2010,ER2010,LBJ2010,ZR2011}.  The second deficiency is that of computational complexity: while Bayesian MMV algorithms appear to offer the strongest recovery performance, it comes at the cost of increased complexity relative to simpler schemes, such as those based on greedy pursuit.  For high-dimensional datasets, the complexity of Bayesian techniques may prohibit their application.}

{Our goal is to develop an MMV algorithm that offers the best of both worlds, combining the recovery performance of Bayesian techniques, even in the presence of substantial amplitude correlation and apriori unknown signal statistics, with the linear complexity scaling of greedy pursuit methods.  Aiding us in meeting our goal is a powerful algorithmic framework known as \emph{approximate message passing} (AMP), first proposed by Donoho et al. for the SMV CS problem \cite{DMM2009}.  In its early SMV formulations, AMP was shown to perform rapid and highly accurate probabilistic inference on models with known i.i.d. signal and noise priors, and i.i.d. random $\vec{A}$ matrices \cite{DMM2009,DMM2010}.  More recently, AMP was extended to the block-sparse SMV problem under similar conditions \cite{DJM2011}.  While this block-sparse SMV AMP does solve a simple version of the MMV problem via the formulation \eqref{smv_linear_model}, it does not account for intra-block amplitude correlation (i.e., temporal correlation in the MMV model).  Recently, Kim et al. proposed an AMP-based MMV algorithm that does exploit temporal amplitude correlation \cite{KCJBY2011}.  However, their approach requires knowledge of the signal and noise statistics (e.g., sparsity, power, correlation) and uses matrix inversions at each iteration, implying a complexity that grows superlinearly in the problem dimensions.}

{In this work, we propose an AMP-based MMV algorithm (henceforth referred to as AMP-MMV) that exploits temporal amplitude correlation and learns the signal and noise statistics directly from the data, all while maintaining a computational complexity that grows linearly in the problem dimensions.  In addition, our AMP-MMV can easily accomodate time-varying measurement matrices $\vect{A}$, implicit measurement operators (e.g., FFT $\vect{A}$), and complex-valued quantities.  (These latter scenarios occur in, e.g., digital communication \cite{BHSN2010} and pMRI \cite{OS2006}.)  The key to our approach lies in combining the ``turbo AMP'' framework of \cite{S2010a}, where the usual AMP factor graph is augmented with additional hidden variable nodes and inference is performed on the augmented factor graph, with an EM-based approach to hyperparameter learning.  Details are provided in Sections \ref{sec:signal_model}, \ref{sec:algorithm}, and \ref{sec:parameter_estimation}.}

{In \secref{numerical_study}, we present a detailed numerical study of AMP-MMV that includes a comparison against three state-of-the-art MMV algorithms.  In order to establish an absolute performance benchmark, in \secref{kalman} we describe a tight, oracle-aided performance lower bound that is realized through a support-aware Kalman smoother (SKS).  To the best of our knowledge, our numerical study is the first in the MMV literature to use the SKS as a benchmark.  Our numerical study demonstrates that AMP-MMV performs near this oracle performance bound under a wide range of problem settings, and that AMP-MMV is especially suitable for application to high-dimensional problems.  In what represents a less-explored direction for the MMV problem, we also explore the effects of measurement matrix time-variation (cf. \cite{TMT2010}).  Our results show that measurement matrix time-variation can significantly improve reconstruction performance and thus we advocate the use of time-varying measurement operators whenever possible.}

\subsection{Notation}
\label{sec:introduction:notation}

Boldfaced lower-case letters, e.g., $\vec{a}$, denote vectors, while boldfaced upper-case letters, e.g., $\vec{A}$, denote matrices.  The letter $t$ is strictly used to index a timestep, $t=1, 2, \ldots, T$, the letter $n$ is strictly used to index the coefficients of a signal, $n = 1,\ldots,N$, and the letter $m$ is strictly used to index the measurements, $m=1,\ldots,M$.  The superscript ${}^{(t)}$ indicates a timestep-dependent quantity, while a superscript without parentheses, such as ${}^k$, indicates a quantity whose value changes according to some algorithmic iteration index $k$.  Subscripted variables such as $\nt{x}$ are used to denote the $n^{th}$ element of the vector $\vect{x}$.  The $m^{th}$ row of the matrix $\vec{A}$ is denoted by $\vec{a}_m^{\textsf{T}}$, and the transpose (conjugate transpose) by $\vec{A}^\textsf{T}$ ($\vec{A}^{\textsf{H}}$).  An $M$-by-$M$ identity matrix is denoted by $\vec{I}_{_{M}}$, a length-$N$ vector of ones is given by $\vec{1}_{_N}$ and $\vec{\mathcal{D}}(\vec{a})$ designates a diagonal matrix whose diagonal entries are given by the elements of the vector $\vec{a}$.  Finally, $\mathcal{CN}(\vec{a};\vec{b},\vec{C})$ refers to the complex normal distribution that is a function of the vector $\vec{a}$, with mean $\vec{b}$ and covariance matrix $\vec{C}$.

\section{Signal Model}
\label{sec:signal_model}
In this section, we elaborate on the signal model outlined in \secref{introduction}, and make precise our modeling assumptions.  Our signal model, as well as our algorithm, will be presented in the context of complex-valued signals, but can be easily modified to accommodate real-valued signals.

As noted in \secref{introduction}, we consider the linear measurement model \eqref{linear_model}, in which the signal $\vect{x} \in \mathbb{C}^N$ at timestep $t$ is observed as $\vect{y} \in \mathbb{C}^M$ through the linear operator $\vec{A} \in \mathbb{C}^{M \times N}$.  We assume $\vect{e} \sim \mathcal{CN}(\vec{0}, \sigma_e^2 \vec{I}_{_{M}})$ is circularly symmetric complex white Gaussian noise.  We use $\mathcal{S} \triangleq \{n | \nt{x} \neq 0\}$ to denote the indices of the time-invariant support of the signal, which is assumed to be suitably sparse, i.e., $|\mathcal{S}| \le M$.\footnote{If the signal being recovered is not itself sparse, it is assumed that there exists a known basis, incoherent with the measurement matrix, in which the signal possesses a sparse representation.  Without loss of generality, we will assume the underlying signal is sparse in the canonical basis.}

Our approach to specifying a prior distribution for the signal, $p(\{\vect{x}\}_{t=1}^T)$, is motivated by a desire to separate the support, $\mathcal{S}$, from the amplitudes of the non-zero, or ``active,'' coefficients.  To accomplish this, we decompose each coefficient $\nt{x}$ as the product of two hidden variables:
\begin{equation}
	\nt{x} = s_n\cdot \nt{\theta} \qquad \Leftrightarrow \qquad p(\nt{x}|s_n,\nt{\theta}) = \left\{
	\begin{array}{ll}
		\delta(\nt{x} - \nt{\theta}), & s_n = 1, \\
		\delta(\nt{x}), & s_n = 0,
	\end{array}
	\right. 
	\label{eq:x_decomp}
\end{equation}
where $s_n \in \{0,1\}$ is a binary variable that indicates support set membership, and $\nt{\theta} \in \mathbb{C}$ is a variable that provides the amplitude of coefficient $\nt{x}$.  When $s_n = 0$, $\nt{x} = 0$ and $n \notin \mathcal{S}$, and when $s_n = 1$, $\nt{x} = \nt{\theta}$ and $n \in \mathcal{S}$.  To model the sparsity of the signal, we treat each $s_n$ as a Bernoulli random variable with $\text{Pr}\{s_n = 1\} \triangleq \lambda_n < 1$.

In order to model the temporal correlation of signal amplitudes, we treat the evolution of amplitudes over time as stationary first-order Gauss-Markov random processes.  Specifically, we assume that $\nt{\theta}$ evolves according to the following linear dynamical system model:
\begin{equation}
	\nt{\theta} = (1 - \alpha)(\theta_n^{(t-1)} - \zeta) + \alpha \nt{w} + \zeta,
	\label{eq:theta_evolve}
\end{equation}
where $\zeta \in \mathbb{C}$ is the mean of the amplitude process, $\nt{w} \sim \mathcal{CN}(0, \rho)$ is a circularly symmetric white Gaussian perturbation process, and $\alpha \in [0, 1]$ is a scalar that controls the correlation of $\nt{\theta}$ across time.  At one extreme, $\alpha = 0$, the random process is perfectly correlated $(\nt{\theta} = \theta_n^{(t-1)})$, while at the other extreme, $\alpha = 1$, the amplitudes evolve independently over time.  Note that the binary support vector, $\vec{s}$, is independent of the amplitude random process, $\{\vect{\theta}\}_{t=1}^T$, which implies that there are hidden amplitude ``trajectories'', $\{\nt{\theta}\}_{t=1}^T$, associated with inactive coefficients.  Consequently, $\nt{\theta}$ should be thought of as the conditional amplitude of $\nt{x}$, conditioned on $s_n = 1$.

Under our model, the prior distribution of any signal coefficient, $\nt{x}$, is a Bernoulli-Gaussian or ``spike-and-slab'' distribution:
\begin{equation}
	p(\nt{x}) = (1 - \lambda_n) \delta\big(\nt{x}\big) + \lambda_n \mathcal{CN}\big(\nt{x}; \zeta, \sigma^2 \big),
	\label{eq:x_prior}
\end{equation}
where $\delta(\cdot)$ is the Dirac delta function and $\sigma^2 \triangleq \tfrac{\alpha \rho}{2 - \alpha}$ is the steady-state variance of $\nt{\theta}$.  We note that when $\lambda_n < 1$, \eqref{x_prior} is an effective sparsity-promoting prior due to the point mass at $\nt{x} = 0$.

\section{The Support-Aware Kalman Smoother}
\label{sec:kalman}
Prior to describing AMP-MMV in detail, we first motivate the type of inference we wish to perform.  Suppose for a moment that we are interested in obtaining a minimum mean square error (MMSE) estimate of $\{\vect{x}\}_{t=1}^T$, and that we have access to an oracle who can provide us with the support, $\mathcal{S}$.  With this knowledge, we can concentrate solely on estimating $\{\vect{\theta}\}_{t=1}^T$, since, conditioned on $\mathcal{S}$, an MMSE estimate of $\{\vect{\theta}\}_{t=1}^T$ can provide an MMSE estimate of $\{\vect{x}\}_{t=1}^T$.  For the linear dynamical system of \eqref{theta_evolve}, the support-aware Kalman smoother (SKS) provides the appropriate oracle-aided MMSE estimator of $\{\vect{\theta}\}_{t=1}^T$ \cite{E2006}.  The state-space model used by the SKS is:
\begin{eqnarray}
	\vect{\theta} &=& (1-\alpha) \vec{\theta}^{(t-1)} + \alpha \zeta \vec{1}_{_{N}} + \alpha \vect{w}, \label{eq:theta_ss_full} \\
	\vect{y} &=& \vec{A} \vec{\mathcal{D}}(\vec{s}) \vect{\theta} + \vect{e}, \label{eq:y_ss_full}
\end{eqnarray}
where $\vec{s}$ is the binary support vector associated with $\mathcal{S}$.  If $\hvec{\theta}^{(t)}$ is the MMSE estimate returned by the SKS, then an MMSE estimate of $\vect{x}$ is given by $\hvec{x}^{(t)} = \vec{\mathcal{D}}(\vec{s}) \hvec{\theta}^{(t)}$.

The state-space model \eqref{theta_ss_full}, \eqref{y_ss_full} provides a useful interpretation of our signal model.  In the context of Kalman smoothing, the state vector $\vect{\theta}$ is only partially observable (due to the action of $\vec{\mathcal{D}}(\vec{s})$ in \eqref{y_ss_full}).  Since $\vec{\mathcal{D}}(\vec{s}) \vect{\theta} = \vect{x}$, noisy linear measurements of $\vect{x}$ are used to infer the state $\vect{\theta}$.  However, since only those $\nt{\theta}$ for which $n \in \mathcal{S}$ are observable, and thus identifiable, they are the only ones whose posterior distributions will be meaningful.

Since the SKS performs optimal MMSE estimation, given knowledge of the true signal support, it provides a useful lower bound on the achievable performance of any support-agnostic Bayesian algorithm that aims to perform MMSE estimation of $\{\vect{x}\}_{t=1}^T$.

\section{The AMP-MMV Algorithm}
\label{sec:algorithm}
In \secref{signal_model}, we decomposed each signal coefficient, $\nt{x}$, as the product of a binary support variable, $s_n$, and an amplitude variable, $\nt{\theta}$.  We now develop an algorithm that infers a marginal posterior distribution on each variable, enabling both soft estimation and soft support detection.

\begin{figure}
	\begin{center}
		\includegraphics[angle=-90,bb=55 250  400 500,scale=0.50]{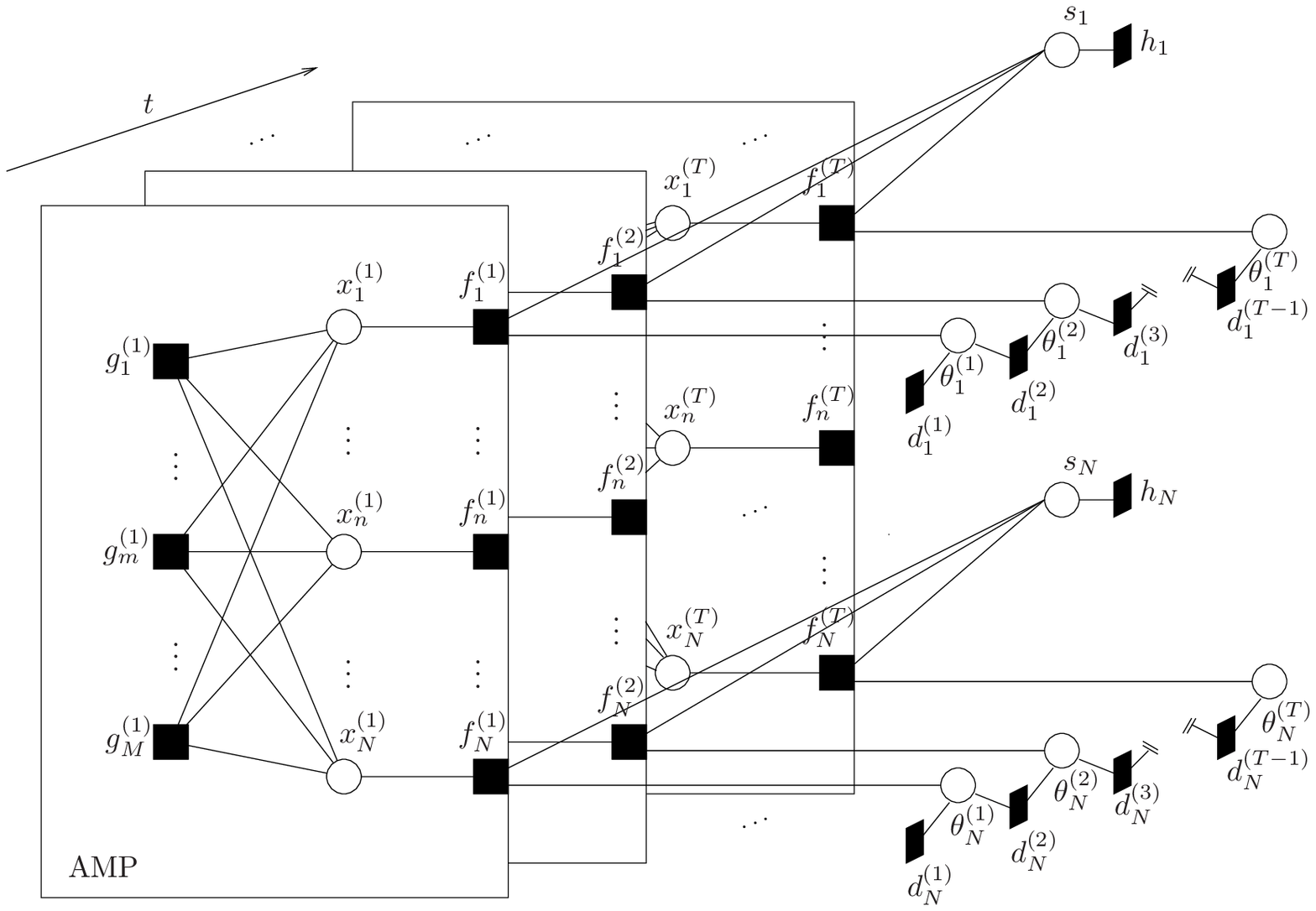}
		\caption{Factor graph representation of the decomposition of $p(\ovec{x},\ovec{\theta},\vec{s}|\ovec{y})$ in \eqref{joint_decomp}.}
		\label{fig:total_factor_graph}
	\end{center}
\end{figure}

The statistical structure of the signal model from \secref{signal_model} becomes apparent from a factorization of the posterior joint pdf of all random variables.  {Recalling from \eqref{smv_linear_model} the definitions of $\ovec{y}$ and $\ovec{x}$, and defining} $\ovec{\theta}$ similarly, the posterior joint distribution factors as follows:
\ifthenelse{\boolean{ONE_COLUMN}}
{
\begin{equation}
	p(\ovec{x}, \ovec{\theta}, \vec{s} | \ovec{y}) \propto \prod_{t=1}^T \left( \prod_{m=1}^M p(y_m^{(t)} | \vect{x}) \prod_{n=1}^N p(\nt{x} | \nt{\theta}, s_n) p(\nt{\theta} | \theta_n^{(t-1)}) \right) \prod_{n=1}^N p(s_n),
	\label{eq:joint_decomp}
\end{equation}
}
{
\begin{eqnarray}
	p(\ovec{x}, \ovec{\theta}, \vec{s} | \ovec{y}) &\propto& \prod_{t=1}^T \Bigg( \prod_{m=1}^M p(y_m^{(t)} | \vect{x}) \prod_{n=1}^N p(\nt{x} | \nt{\theta}, s_n)	\nonumber \\
	&\quad& \qquad \times \, p(\nt{\theta} | \theta_n^{(t-1)}) \Bigg) \prod_{n=1}^N p(s_n),
	\label{eq:joint_decomp}
\end{eqnarray}
}
where $\propto$ indicates equality up to a normalizing constant, and $p(\theta_n^{(1)} | \theta_n^{(0)}) \triangleq p(\theta_n^{(1)})$.  A convenient graphical representation of this decomposition is given by a \emph{factor graph} \cite{KFL2001}, which is an undirected bipartite graph that connects the pdf ``factors'' of \eqref{joint_decomp} with the variables that make up their arguments.  The factor graph for the decomposition of \eqref{joint_decomp} is shown in \figref{total_factor_graph}.  The \emph{factor nodes} are denoted by filled squares, while the \emph{variable nodes} are denoted by circles.  In the figure, the signal variable nodes at timestep $t$, $\{\nt{x}\}_{n=1}^N$, are depicted as lying in a plane, or ``frame'', with successive frames stacked one after another.  Since during inference the measurements $\{y_m^{(t)}\}$ are known observations and not random variables, they do not appear explicitly in the factor graph.  The connection between the frames occurs through the amplitude and support indicator variables, providing a graphical representation of the temporal correlation in the signal.  For visual clarity, these $\{\nt{\theta}\}_{t=1}^T$ and $s_n$ variable nodes have been removed from the graph for the intermediate index $n$, but should in fact be present at every index $n = 1,\ldots,N$.

The factor nodes in \figref{total_factor_graph} have all been assigned alphabetic labels; the correspondence between these labels and the distributions they represent, as well as the functional form of each distribution, is presented in \tabref{factors}.
\begin{table*}[t]
	\begin{center}
	\begin{tabular}{rcl}
		Factor	&	Distribution		&	Functional Form \\
		\hline
		$g_m^{(t)}\big(\vect{x}\big)$	&	$p\big(y_m^{(t)} | \vec{x}^{(t)}\big)$	&	$\mathcal{CN}\big(y_m^{(t)}; \vec{a}_m^{\textsf{T}} \vec{x}^{(t)}, \sigma_e^2\big)$	\\
		$\nt{f}\big(\nt{x}, s_n, \nt{\theta}\big)$	&	$p\big(\nt{x} | s_n, \nt{\theta}\big)$	&	$\delta \big(\nt{x} - s_n \nt{\theta}\big)$	\\
		$h_n\big(s_n\big)$	&	$p\big(s_n\big)$	&	$\big(1 - \lambda_{n}\big)^{(1 - s_n)} \big(\lambda_{n}\big)^{s_n}$	\\
		$d_n^{(1)}\big(\theta_n^{(1)}\big)$	&	$p\big(\theta_n^{(1)}\big)$	&	$\mathcal{CN}\big(\theta_n^{(1)}; \zeta, \sigma^2\big)$	\\
		$\nt{d}\big(\nt{\theta}, \theta_n^{(t-1)}\big)$	&	$p\big(\nt{\theta} | \theta_n^{(t-1)}\big)$	&	$\mathcal{CN}\big(\nt{\theta}; (1 - \alpha) \theta_n^{(t-1)} + \alpha \zeta, \alpha^2 \rho \big)$	\\
	\end{tabular}
	\caption{The factors, underlying distributions, and functional forms associated with the signal model of \secref{signal_model}.}
	\label{tab:factors}
	\end{center}
\end{table*}

A natural approach to performing statistical inference on a signal model that possesses a convenient factor graph representation is through a message passing algorithm known as belief propagation \cite{P1988}.  In belief propagation, the messages exchanged between connected nodes of the graph represent probability distributions.  In cycle-free graphs, belief propagation can be viewed as an instance of the sum-product algorithm \cite{KFL2001}, allowing one to obtain an exact posterior marginal distribution for each unobserved variable, given a collection of observed variables.  When the factor graph contains cycles, the same rules that define the sum-product algorithm can still be applied, however convergence is no longer guaranteed \cite{KFL2001}.  Despite this, there exist many problems to which loopy belief propagation \cite{FM1998} has been successfully applied, including inference on Markov random fields \cite{FPC2000}, LDPC decoding \cite{M2003}, and compressed sensing \cite{DMM2009, BSB2010, S2010a, 
R2010a, R2010b, 
BM2011}.

We now proceed with a high-level description of AMP-MMV, an algorithm that follows the sum-product methodology while leveraging recent advances in message approximation \cite{DMM2009}.  In what follows, we use $\msg{a}{b}(\cdot)$ to denote a message that is passed from node $a$ to a connected node $b$.

\subsection{Message Scheduling}
\label{sec:scheduling}
Since the factor graph of \figref{total_factor_graph} contains many cycles there are a number of valid ways to schedule, or sequence,  the messages that are exchanged in the graph.  We will describe two message passing schedules that empirically provide good convergence behavior and straightforward implementation.  We refer to these two schedules as the \emph{parallel message schedule} and the \emph{serial message schedule}.  In both cases, messages are first initialized to agnostic values, and then iteratively exchanged throughout the graph according to the chosen schedule until either convergence occurs, or a maximum number of allowable iterations is reached.

Conceptually, both message schedules can be decomposed into four distinct phases, differing only in which messages are initialized and the order in which the phases are sequenced.  We label each phase using the mnemonics \textbf{(into)}, \textbf{(within)}, \textbf{(out)}, and \textbf{(across)}.  In phase \textbf{(into)}, messages are passed from the $s_n$ and $\nt{\theta}$ variable nodes \emph{into} frame $t$.  Loosely speaking, these messages convey current beliefs about the values of $\vec{s}$ and $\vect{\theta}$.  In phase \textbf{(within)}, messages are exchanged \emph{within} frame $t$, producing an estimate of $\vect{x}$ using the current beliefs about $\vec{s}$ and $\vect{\theta}$ together with the available measurements $\vect{y}$.  In phase \textbf{(out)}, the estimate of $\vect{x}$ is used to refine the beliefs about $\vec{s}$ and $\vect{\theta}$ by passing messages \emph{out} of frame $t$.  Finally, in phase \textbf{(across)}, messages are sent from $\nt{\theta}$ to either $\theta_n^{(t+1)}$ or $\theta_n^{(t-1)}$, thus conveying information \emph{across} time about temporal correlation in the signal amplitudes.

The parallel message schedule begins by performing phase \textbf{(into)} in parallel for each frame $t = 1, \ldots, T$ simultaneously.  Then, phase \textbf{(within)} is performed simultaneously for each frame, followed by phase \textbf{(out)}.  Next, information about the amplitudes is exchanged between the different timesteps by performing phase \textbf{(across)} in the forward direction, i.e., messages are passed from $\theta_n^{(1)}$ to $\theta_n^{(2)}$, and then from $\theta_n^{(2)}$ to $\theta_n^{(3)}$, proceeding until $\theta_n^{(T)}$ is reached.  Finally, phase \textbf{(across)} is performed in the backward direction, where messages are passed consecutively from $\theta_n^{(T)}$ down to $\theta_n^{(1)}$.  At this point, a single iteration of AMP-MMV has been completed, and a new iteration can commence starting with phase \textbf{(into)}.  In this way, all of the available measurements, $\{\vect{y}\}_{t=1}^T$, are used to influence the recovery of the signal at each timestep.

The serial message schedule is similar to the parallel schedule except that it operates on frames in a sequential fashion, enabling causal processing of MMV signals.  Beginning at the initial timestep, $t = 1$, the serial schedule first performs phase \textbf{(into)}, followed by phases \textbf{(within)} and \textbf{(out)}.  Outgoing messages from the initial frame are then used in phase \textbf{(across)} to pass messages from $\theta_n^{(1)}$ to $\theta_n^{(2)}$.  The messages arriving at $\theta_n^{(2)}$, along with updated beliefs about the value of $\vec{s}$, are used to initiate phase \textbf{(into)} at timestep $t = 2$.  Phases \textbf{(within)} and \textbf{(out)} are performed for frame $2$, followed by another round of phase \textbf{(across)}, with messages being passed forward to $\theta_n^{(3)}$.  This procedure continues until phase \textbf{(out)} is completed at frame $T$.  Until now, only causal information has been used in producing estimates of the signal.  If the application permits smoothing, then message passing continues in a similar fashion, but with messages now propagating backward in time, i.e., messages are passed from $\theta_n^{(T)}$ to $\theta_n^{(T-1)}$, phases \textbf{(into)}, \textbf{(within)}, and \textbf{(out)} are performed at frame $T-1$, and then messages move from $\theta_n^{(T-1)}$ to $\theta_n^{(T-2)}$.  The process continues until messages arrive at $\theta_n^{(1)}$, at which point a single \emph{forward/backward pass} has been completed.  We complete multiple such passes, resulting in a smoothed estimate of the signal.

\subsection{Implementing the Message Passes}
\label{sec:implementation}
Space constraints prohibit us from providing a full derivation of all the messages that are exchanged through the factor graph of \figref{total_factor_graph}.  Most messages can be derived by straightforward application of the rules of the sum-product algorithm.  Therefore, in this sub-section we will restrict our attention to a handful of messages in the \textbf{(within)} and \textbf{(out)} phases whose implementation requires a departure from the sum-product rules for one reason or another.

\begin{figure}
	\begin{center}
		\scalebox{0.55}{\rotatebox{90}{\includegraphics*[1.50in,2.75in][6.55in,9.65in]{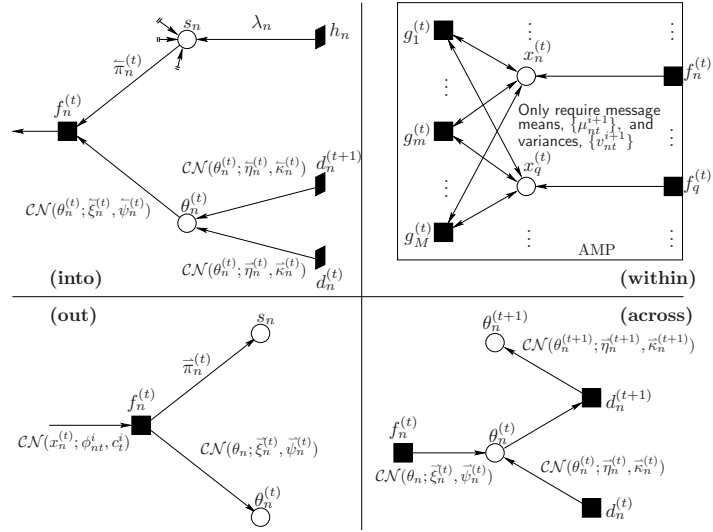}}}
		\caption{A summary of the four message passing phases, including message notation and form.}
		\label{fig:message_summary}
	\end{center}
\end{figure}

To aid our discussion, in \figref{message_summary} we summarize each of the four phases, focusing primarily on a single coefficient index $n$ at some intermediate frame $t$.  Arrows indicate the direction that messages are moving, and only those nodes and edges participating in a particular phase are shown in that phase.  For the \textbf{(across)} phase we show messages being passed forward in time, and omit a graphic for the corresponding backwards pass.  The figure also introduces the notation that we adopt for the different variables that serve to parameterize the messages.  Certain variables, e.g., $\nt{\fwd{\eta}}$ and $\nt{\bwd{\eta}}$, are accented with directional arrows.  This is to distinguish variables associated with messages moving in one direction from those associated with messages moving in another.  For Bernoulli message pdfs, we show only the nonzero probability, e.g., $\lambda_n = \nu_{h_n \to s_n}(s_n =1)$.

Phase \textbf{(within)} entails using the messages transmitted from $s_n$ and $\nt{\theta}$ to $\nt{f}$ to compute the messages that pass between $\nt{x}$ and the $\{g_m^{(t)}\}$ nodes.  Inspection of \figref{message_summary} reveals a dense interconnection between the $\{\nt{x}\}$ and $\{g_m^{(t)}\}$ nodes.  As a consequence, applying the standard sum-product rules to compute the  $\msg{g_m^{(t)}}{\nt{x}}(\cdot)$ messages would result in an algorithm that required the evaluation of multi-dimensional integrals that grew exponentially in number in both $N$ and $M$.  Since we are strongly motivated to apply AMP-MMV to high-dimensional problems, this approach is clearly infeasible.  Instead, we turn to a recently developed algorithm known as \emph{approximate message passing} (AMP).

{AMP was originally proposed by Donoho et al. \cite{DMM2009} as a message passing algorithm designed to solve the noiseless SMV CS problem known as Basis Pursuit ($\min \|\vec{x}\|_1 \text{ s.t. } \vec{y} = \vec{Ax}$), and was subsequently extended \cite{DMM2010} to support MMSE estimation under white-Gaussian-noise-corrupted observations and generic signal priors of the form $p(\vec{x}) = \prod p(x_n)$ through an approximation of the sum-product algorithm.  In both cases, the associated factor graph looks identical to that of the \textbf{(within)} segment of \figref{message_summary}.  Conventional wisdom holds that loopy belief propagation only works well when the factor graph is locally tree-like.  For general, non-sparse $\vec{A}$ matrices, the \textbf{(within)} graph will clearly not possess this property, due to the many short cycles between the $\nt{x}$ and $g_m^{(t)}$ nodes.  Reasoning differently, Donoho et al. showed that the density of connections could prove beneficial, if properly exploited.
}

{In particular, central limit theorem arguments suggest that the messages propagated from the $g_m$ nodes to the $x_n$ nodes under the sum-product algorithm can be well-approximated as Gaussian when the problem dimensionality is sufficiently high.  Moreover, the computation of these Gaussian-approximated messages only requires knowledge of the mean and variance of the sum-product messages from the $x_n$ to the $g_m$ nodes.  Finally, when $|A_{mn}|^2$ scales as $\mathcal{O}(1/M)$ for all $(m,n)$, the differences between the variances of the messages emitted by the $x_n$ nodes vanish as $M$ grows large, as do those of the $g_m$ nodes when $N$ grows large, allowing each to be approximated by a single, common variance.  Together, these sum-product approximations yield an iterative thresholding algorithm with a particular first-order correction term that ensures both Gaussianity and independence in the residual error vector over the iterations.  The complexity of this iterative thresholding algorithm is dominated by a single multiplication by $\vec{A}$ and $\vec{A}^{\textsf{H}}$ per iteration, implying a per-iteration computational cost of $\mathcal{O}(MN)$ flops.  Furthermore, the state-evolution equation that governs the transient behavior of AMP shows that the number of required iterations does not scale with either $M$ or $N$, implying that the total complexity is itself $\mathcal{O}(MN)$ flops.}

{AMP's suitability for the MMV problem stems from several considerations.  First, AMP's probabilistic construction, coupled with its message passing implementation, makes it well-suited for incorporation as a subroutine within a larger message passing algorithm.  In the MMV problem it is clear that $p(\ovec{x}) \neq \prod p(\nt{x})$ due to the joint sparsity and amplitude correlation structure, and therefore AMP does not appear to be directly applicable.  Fortunately, by modeling this structure through the hidden variables $\vec{s}$ and $\ovec{\theta}$, we can exploit the conditional independence of the signal coefficients: $p(\ovec{x} | \vec{s}, \ovec{\theta}) = \prod p(\nt{x} | s_n, \nt{\theta})$.  In particular, we replace the $p(\nt{x})$ that AMP traditionally expects with $\msg{\nt{f}}{\nt{x}}(\cdot)$, the most recent message moving into the \textbf{(within)} segment of \figref{message_summary}.  This message represents a ``local prior'' on $\nt{x}$ given the current belief about the hidden variables $s_n$ and $\nt{\theta}$, and assumes the Bernoulli-Gaussian form
\begin{equation}
	\nu_{f_n^{(t)} \to x_n^{(t)}}(x_n^{(t)}) = (1 - \nt{\bwd{\pi}}) \delta(x_n^{(t)}) + \nt{\bwd{\pi}} \mathcal{CN}(x_n^{(t)}; \bwd{\xi}_n^{(t)}, \bwd{\psi}_n^{(t)}).
	\label{eq:f_to_x_msg}
\end{equation}
This ``local prior'' determines the AMP soft-thresholding functions defined in \eqref{thresh_start} - \eqref{thresh_end} of \tabref{algorithm_equations}.  The derivation of these thresholding functions closely follows those outlined in \cite{S2010a}, which considered the special case of a zero-mean Bernoulli-Gaussian prior.}

{Beyond the ease with which AMP is included into the larger message passing algorithm, a second factor that favors using AMP is the tremendous computational efficiency it imparts on high-dimensional problems.  Using AMP to perform the most computationally intensive message passes enables AMP-MMV to attain a linear complexity scaling in all problem dimensions.  To see why this is the case, note that the \textbf{(into)}, \textbf{(out)}, and \textbf{(across)} steps can be executed in $\mathcal{O}(N)$ flops/timestep, while AMP allows the \textbf{(within)} step to be executed in $\mathcal{O}(MN)$ flops/timestep (see \eqref{amp_start} - \eqref{amp_end} of \tabref{algorithm_equations}).  Since these four steps are executed $\mathcal{O}(T)$ times per AMP-MMV iteration for both the serial and parallel message schedules, it follows that AMP-MMV's overall complexity is $\mathcal{O}(TMN)$.\footnote{{The primary computational burden of executing AMP-MMV involves performing matrix-vector products with $\vec{A}$ and $\vec{A}^\textsf{H}$, allowing it to be easily applied in problems where the measurement matrix is never stored explicitly, but rather is implemented implicitly through subroutines.}  Fast implicit $\vec{A}$ operators can provide significant computational savings in high-dimensional problems; implementing a Fourier transform as a fast Fourier transform (FFT) subroutine, for example, would drop AMP-MMV's complexity from $\mathcal{O}(TMN)$ to $\mathcal{O}(TN \log_2 N)$.}}

{A third appealing feature of AMP is that it is theoretically well-grounded; a recent analysis \cite{BM2011} shows that, for Gaussian $\vec{A}$ in the large-system limit (i.e., $M$, $N \to \infty$ with $M$/$N$ fixed), the behavior of AMP is governed by a state evolution whose fixed points, when unique, correspond to MMSE-optimal signal estimates.}

\begin{table}[t]
\centering
\scriptsize
\setlength{\tabcolsep}{2pt}
\setlength{\belowcaptionskip}{3ex}
\begin{tabular}{|llrclr|}
	\hline
	\multicolumn{5}{|l}{$\textsf{\% Define soft-thresholding functions: }$}&\\
	& \multicolumn{4}{l}{$\textit{F}_{nt}(\phi; c) \triangleq (1 + \gamma_{nt}(\phi; c))^{-1}  
		\Big( \frac{\bwd{\psi}_n^{(t)} \phi + \bwd{\xi}_n^{(t)} c}{\bwd{\psi}_n^{(t)} + c} \Big)$}	& \threshcnt{eq:thresh_start}\\[2ex]
	& \multicolumn{4}{l}{$\textit{G}_{nt}(\phi; c) \triangleq (1 + \gamma_{nt}(\phi; c))^{-1} 
		\Big( \frac{\bwd{\psi}_n^{(t)} c}{\bwd{\psi}_n^{(t)} + c} \Big) + \gamma_{nt}(\phi; c) |\textit{F}_n(\phi; c)|^2$}	& \threshcnt{}\\[1ex]
	& \multicolumn{4}{l}{$\textit{F}_{nt}'(\phi; c) \triangleq \tfrac{\partial}{\partial \phi} \textit{F}_{nt}(\phi,c) = \tfrac{1}{c} \textit{G}_{nt}(\phi; c)$}	& \threshcnt{}\\[1ex]
	& \multicolumn{4}{l}{$\gamma_{nt}(\phi; c) \triangleq \Big( \frac{1 - \bwd{\pi}_n^{(t)}}{\bwd{\pi}_n^{(t)}} \Big) \Big( \frac{\bwd{\psi}_n^{(t)} + c}{c} \Big) $} &\\
	&&& \multicolumn{2}{l}{$\quad \times \exp\Big( - \Big[ \frac{\bwd{\psi}_n^{(t)} |\phi|^2 + \bwd{\xi}_n^{(t)\,*} c \phi + 
		\bwd{\xi}_n^{(t)} c \phi^* - c |\bwd{\xi}_n^{(t)}|^2}{c(\bwd{\psi}_n^{(t)} + c)} \Big] \Big)$}	& \threshcnt{eq:thresh_end}\\[1ex]
	\hline
	\multicolumn{5}{|l}{$\textsf{\% Begin passing messages} \ldots$} &\\[-1ex]
  	\multicolumn{5}{|l}{$\textsf{for } t=1,\ldots,T, \, \forall n:$}&\\[-1ex]
	&\multicolumn{4}{l}{$\quad \textsf{\% Execute the } \textbf{(into)} \textsf{ phase} \ldots$} &\\
	&\multicolumn{4}{l}{$\quad \nt{\bwd{\pi}} = \frac {\lambda_n \cdot \prod_{t' \neq t} \fwd{\pi}_n^{(t')}} 
		{(1 - \lambda_n) \cdot \prod_{t' \neq t} (1 - \fwd{\pi}_n^{(t')}) + \lambda_n \cdot \prod_{t' \neq t} \fwd{\pi}_n^{(t')}} $}	& \algcnt{} \\[2mm]
	&\multicolumn{4}{l}{$\quad \bwd{\psi}_{n}^{(t)} = \frac{\fwd{\kappa}_{n}^{(t)} \cdot \bwd{\kappa}_{n}^{(t)}}
		{\fwd{\kappa}_{n}^{(t)} + \bwd{\kappa}_{n}^{(t)}}$}	& \algcnt{}\\[2mm]
	&\multicolumn{4}{l}{$\quad \bwd{\xi}_{n}^{(t)} = \bwd{\psi}_{n}^{(t)} \cdot \Big(\frac{\fwd{\eta}_{n}^{(t)}}
		{\fwd{\kappa}_{n}^{(t)}} + \frac{\bwd{\eta}_{n}^{(t)}}{\bwd{\kappa}_{n}^{(t)}}\Big)$}	& \algcnt{}\\

	&\multicolumn{4}{l}{$\quad \textsf{\% Initialize AMP-related variables} \ldots$} &\\
	&\multicolumn{4}{l}{$\quad \forall m: z_{mt}^1 = y_{m}^{(t)}, \forall n: \mu_{nt}^1 = 0, \textsf{ and } c_t^1 = 100 \cdot \textstyle \sum_{n=1}^N \psi_n^{(t)}$}&\\

	&\multicolumn{5}{l|}{$\quad \textsf{\% Execute the } \textbf{(within)} \textsf{ phase using AMP} \ldots$} \\
  	&\multicolumn{4}{l}{$\quad \textsf{for $i=1,\ldots,I$, } \forall n,m:$}&\\
  	&&$\qquad \phi_{nt}^i$ &$=$& $\sum_{m=1}^M A_{mn}^{*} z_{mt}^i + \mu_{nt}^i$ & \algcnt{eq:amp_start}\\
	&&$\qquad \mu_{nt}^{i+1}$ &$=$& $\textit{F}_{nt}(\phi_{nt}^i; c_t^i)$	&\algcnt{eq:amp_mu_defn}\\
	&&$\qquad v_{nt}^{i+1}$ &$=$& $\textit{G}_{nt}(\phi_{nt}^i; c_t^i)$	& \algcnt{eq:amp_v_defn}\\
	&&$\qquad c_t^{i+1}$ &$=$& $\sigma_e^2 + \tfrac{1}{M} \sum_{n=1}^N v_{nt}^{i+1}$	& \algcnt{}\\
	&&$\qquad z_{mt}^{i+1}$ &$=$& $y_m^{(t)} - \vec{a}_{m}^{\textsf{T}} \vec{\mu}_t^{i+1} + \tfrac{z_{mt}^i}{M} \sum_{n=1}^N \textit{F}_{nt}'(\phi_{nt}^i; c_t^i)$	& \algcnt{eq:amp_end}\\
  	&\multicolumn{2}{l}{$\quad \textsf{end}$}&&&\\
	&\multicolumn{4}{l}{$\quad \hat{x}_n^{(t)} = \mu_{nt}^{I+1} \qquad \textsf{\% Store current estimate of } x_n^{(t)}$}	& \algcnt{eq:amp_x_hat}\\

	&\multicolumn{5}{l|}{$\quad \textsf{\% Execute the } \textbf{(out)} \textsf{ phase} \ldots$} \\
	&\multicolumn{4}{l}{$\quad \fwd{\pi}_n^{(t)} = \Big(1 + \Big( \frac{\bwd{\pi}_n^{(t)}}{1 - \bwd{\pi}_n^{(t)}} \Big) \gamma_{nt}(\phi_{nt}^{I}; c_{t}^{I+1}) \Big)^{-1}$}	& \algcnt{}\\[0ex]
	&\multicolumn{4}{l}{$\quad (\nt{\fwd{\xi}}, \nt{\fwd{\psi}}) = \textsf{taylor\_approx}(\nt{\bwd{\pi}}, \phi_{nt}^I, c_t^I)$}	& \algcnt{}\\[0ex]

	&\multicolumn{5}{l|}{$\quad \textsf{\% Execute the } \textbf{(across)} \textsf{ phase from $\nt{\theta}$ to $\theta_n^{(t+1)}$} \ldots$} \\
	&\multicolumn{4}{l}{$\quad \fwd{\eta}_n^{(t+1)} = (1-\alpha) \Big(\frac{\fwd{\kappa}_{n}^{(t)} \fwd{\psi}_{n}^{(t)}}{\fwd{\kappa}_{n}^{(t)} + \fwd{\psi}_{n}^{(t)}}\Big)
		\Big(\frac{\fwd{\eta}_{n}^{(t)}}{\fwd{\kappa}_{n}^{(t)}} + \frac{\fwd{\xi}_{n}^{(t)}}{\fwd{\psi}_{n}^{(t)}}\Big) + \alpha \zeta$}	& \algcnt{}\\[2ex]
	&\multicolumn{4}{l}{$\quad \fwd{\kappa}_{n}^{(t+1)} = (1-\alpha)^2 \Big(\frac{\fwd{\kappa}_{n}^{(t)} \fwd{\psi}_{n}^{(t)}}{\fwd{\kappa}_{n}^{(t)} + \fwd{\psi}_{n}^{(t)}}\Big) + \alpha^2 \rho$}	& \algcnt{}\\
	\multicolumn{5}{|l}{$\textsf{end}$}&\\
	\hline

\end{tabular}

\caption{Message update equations for executing a single forward pass using the serial message schedule.}
\label{tab:algorithm_equations}
\end{table}

After using AMP to implement phase \textbf{(within)}, we must pass messages out of frame $t$ in order to update our beliefs about the values of $\vec{s}$ and $\vect{\theta}$ in the \textbf{(out)} phase.  Applying the sum-product algorithm rules to compute the message $\msg{\nt{f}}{\nt{\theta}}(\cdot)$ results in the expression
\begin{equation}
	\msg{\nt{f}}{\nt{\theta}}^{\text{exact}}(\nt{\theta}) \triangleq (1 - \nt{\bwd{\pi}}) \mathcal{CN}(0; \phi_{nt}, c_t) + \nt{\bwd{\pi}} \mathcal{CN}(\nt{\theta}; \phi_{nt}, c_t),
	\label{eq:f_to_theta_msg_exact}
\end{equation}
which is an improper distribution due to the constant (w.r.t. $\nt{\theta}$) term $\mathcal{CN}(0; \phi_{nt}, c_t)$.  This behavior is a consequence of the conditional signal model \eqref{x_decomp}.  In particular, when $s_n = 0$, $\nt{x}$ provides no information about the value of $\nt{\theta}$.  Roughly speaking, the term $\mathcal{CN}(0; \phi_{nt}, c_t)$ corresponds to the distribution of $\nt{\theta}$ conditioned on the case $s_n = 0$.

As a means of circumventing the improper message pdf above, we will regard our original signal model, in which $s_n \in \{0,1\}$, as the limiting case of a signal model in which $s_n \in \{\eps,1\}$ with $\eps \to 0$.  For any fixed, positive $\eps$, $\msg{\nt{f}}{\nt{\theta}}(\cdot)$ is given by the proper pdf
\ifthenelse{\boolean{ONE_COLUMN}}
{
\begin{equation}
	\msg{\nt{f}}{\nt{\theta}}^{\text{mod}}(\nt{\theta}) = (1 - \Omega(\nt{\bwd{\pi}})) \,\, \mathcal{CN}(\nt{\theta}; \tfrac{1}{\eps} \phi_{nt}, \tfrac{1}{\eps^2} c_t) + \Omega(\nt{\bwd{\pi}}) \,\, \mathcal{CN}(\nt{\theta}; \phi_{nt}, c_t),
	\label{eq:f_to_theta_msg_mod}
\end{equation}
}
{
\begin{eqnarray}
	\msg{\nt{f}}{\nt{\theta}}^{\text{mod}}(\nt{\theta}) &=& (1 - \Omega(\nt{\bwd{\pi}})) \,\, \mathcal{CN}(\nt{\theta}; \tfrac{1}{\eps} \phi_{nt}, \tfrac{1}{\eps^2} c_t) \nonumber \\
	&\quad& + \,\, \Omega(\nt{\bwd{\pi}}) \,\, \mathcal{CN}(\nt{\theta}; \phi_{nt}, c_t),
	\label{eq:f_to_theta_msg_mod}
\end{eqnarray}
}
where
\begin{equation}
	\Omega(\pi) \triangleq \frac{\eps^2 \pi}{(1 - \pi) + \eps^2 \pi}.
	\label{eq:omega_defn}
\end{equation}
\Eqref{f_to_theta_msg_mod} is a binary Gaussian mixture density.  When $\eps \ll 1$, the first Gaussian component is extremely broad, and conveys little information about the possible value of $\nt{\theta}$.  The second component is a more informative Gaussian whose mean, $\phi_{nt}$, and variance, $c_t$, are determined by the product of the messages $\big\{ \msg{g_m^{(t)}}{\nt{x}}(\cdot) \big\}_{m=1}^M$.  The relative mass assigned to each Gaussian component is a function of the incoming activity probability $\nt{\bwd{\pi}}$ (see \eqref{f_to_x_msg}).  Note that the limiting case of $\Omega(\cdot)$ is a simple indicator function:
\begin{equation}
	\lim_{\eps \to 0} \Omega(\pi) = 
	\begin{cases}
		0 & \text{if } 0 \le \pi < 1, \\
		1 & \text{if } \pi = 1.
	\end{cases}
	\label{eq:omega_indicator}
\end{equation}

When implementing AMP-MMV, we therefore fix $\eps$ at a small positive value, e.g., $\eps = 1 \times 10^{-7}$.  If desired, \eqref{f_to_theta_msg_mod} could then be used as the outgoing message, however this would present a further difficulty.  Propagating a Gaussian mixture along a given edge would result in an exponential growth in the number of mixture components that would need to be propagated along subsequent edges.  To avoid this outcome, we collapse our binary Gaussian mixture to a single Gaussian component, an approach sometimes referred to as \emph{Gaussian sum approximation} \cite{AS1972,BC2010}.  Since, for $\eps \ll 1$, $\Omega(\cdot)$ behaves nearly like the indicator function in \eqref{omega_indicator}, one of the two Gaussian components will typically have negligible mass.  For this reason, collapsing the mixture to a single Gaussian appears justifiable.

To carry out the collapsing, we perform a second-order Taylor series approximation of $- \log \msg{\nt{f}}{\nt{\theta}}^{\text{mod}}(\nt{\theta})$ with respect to $\nt{\theta}$ about the point $\phi_{nt}$.\footnote{For technical reasons, the Taylor series approximation is performed in $\mathbb{R}^2$ instead of $\mathbb{C}$.}  This provides the mean, $\nt{\fwd{\xi}}$, and variance, $\nt{\fwd{\psi}}$, of the single Gaussian that serves as $\msg{\nt{f}}{\nt{\theta}}(\cdot)$.  (See \figref{message_summary}.)  In \appref{taylor} we summarize the Taylor approximation procedure, and in \tabref{taylor_eqs} provide the pseudocode function, \textsf{taylor\_approx}, for computing $\nt{\fwd{\xi}}$ and $\nt{\fwd{\psi}}$.

\putTable{taylor_eqs}{Pseudocode function for computing a single-Gaussian approximation of \eqref{f_to_theta_msg_mod}.}{
\scriptsize
\setlength{\belowcaptionskip}{2ex}
\begin{equation*}
\begin{array}{| llr |}
	\hline
	\multicolumn{3}{| l |}{\textsf{function ($\fwd{\xi}$, $\fwd{\psi}$)  = taylor\_approx($\bwd{\pi}$, $\phi$, $c$)}}\\
	\hline
	
	\multicolumn{3}{| l |}{\textsf{\% Define useful variables:}}\\
	\mf{a} \triangleq \eps^2 (1 - \Omega(\bwd{\pi})) & & \text{(T1)} \\[0ex]
	\bar{\mf{a}} \triangleq \Omega(\bwd{\pi}) & & \text{(T2)} \\[0ex]
	\mf{b} \triangleq \tfrac{\eps^2}{c} |(1 - \tfrac{1}{\eps}) \phi|^2 & & \text{(T3)} \\[0ex]
	\mf{d}_r \triangleq - \tfrac{2 \eps^2}{c} (1 - \tfrac{1}{\eps}) \phi_r & & \text{(T4)} \\[0ex]
	\mf{d}_i \triangleq - \tfrac{2 \eps^2}{c} (1 - \tfrac{1}{\eps}) \phi_i & & \text{(T5)} \\[0ex]
	
	\multicolumn{3}{| l |}{\textsf{\% Compute outputs:}}\\
	\fwd{\psi} = \frac{(\mf{a}^2 e^{-\mf{b}} + \mf{a}\bar{\mf{a}} + \bar{\mf{a}}^2 e^{\mf{b}}) c}{\eps^2 \mf{a}^2 e^{-\mf{b}} + \mf{a}\bar{\mf{a}}(\eps^2 + 1 - \tfrac{1}{2} c \mf{d}_r^2) + \bar{\mf{a}}^2 e^{\mf{b}} } & & \text{(T6)} \\[1ex]
	\fwd{\xi}_r = \phi_r - \tfrac{1}{2} \fwd{\psi} \frac{-\mf{a} e^{-\mf{b}} \mf{d}_r}{\mf{a} e^{-\mf{b}} + \bar{\mf{a}}} & & \text{(T7)} \\[1ex]
	\fwd{\xi}_i = \phi_i - \tfrac{1}{2} \fwd{\psi} \frac{-\mf{a} e^{-\mf{b}} \mf{d}_i}{\mf{a} e^{-\mf{b}} + \bar{\mf{a}}} & & \text{(T8)} \\[1ex]
	\fwd{\xi} = \fwd{\xi}_r + j \fwd{\xi}_i & & \text{(T9)} \\[0ex]
	
	\multicolumn{3}{| l |}{\textsf{return ($\fwd{\xi}$, $\fwd{\psi}$)}}\\
	\hline
\end{array}
\end{equation*}
}

With the exception of the messages discussed above, all the remaining messages can be derived using the standard sum-product algorithm rules \cite{KFL2001}.  For convenience, we summarize the results in \tabref{algorithm_equations}, where we provide a pseudocode implementation of a single forward pass of AMP-MMV using the serial message schedule.

\section{Estimating the Model Parameters}
\label{sec:parameter_estimation}
The signal model of \secref{signal_model} depends on the {sparsity} parameters $\{\lambda_n\}_{n=1}^N$, {amplitude parameters} $\zeta$, $\alpha$, and $\rho$, and {noise variance} $\sigma_e^2$.  While some of these parameters may be known accurately from prior information, it is likely that many will require tuning.  To this end, we develop an expectation-maximization (EM) algorithm that couples with the message passing procedure described in \secref{scheduling} to provide a means of learning all of the model parameters while simultaneously estimating the signal $\ovec{x}$ and its support $\vec{s}$.

The EM algorithm \cite{DLR1977} is an appealing choice for performing parameter estimation for two primary reasons.  First and foremost, the EM algorithm is a well-studied and principled means of parameter estimation.  At every EM iteration, the data likelihood function is guaranteed to increase until convergence to a local maximum of the likelihood function occurs \cite{DLR1977}.  For multimodal likelihood functions, local maxima will, in general, not coincide with the global maximum likelihood (ML) estimator, however a judicious initialization can help in ensuring the EM algorithm reaches the global maximum \cite{M1996}.  {Second, the expectation step of the EM algorithm relies on quantities that have already been computed in the process of executing AMP-MMV.  Ordinarily, this step constitutes the major computational burden of any EM algorithm, thus the fact that we can perform it essentially for free makes our EM procedure highly efficient.}

We let $\Gamma \triangleq \{\lambda, \zeta, \alpha, \rho, \sigma_e^2\}$ denote the set of all model parameters, and let $\Gamma^k$ denote the set of parameter estimates at the $k^{th}$ EM iteration.  Here we have assumed that the binary support indicator variables share a common activity probability, $\lambda$, i.e., $\text{Pr}\{s_n = 1\} = \lambda \,\, \forall n$.  For all parameters except $\sigma_e^2$ we use $\vec{s}$ and $\ovec{\theta}$ as the so-called ``missing'' data of the EM algorithm, while for $\sigma_e^2$ we use $\ovec{x}$.

For the first iteration of AMP-MMV, the model parameters are initialized based on either prior signal knowledge, or according to some heuristic criteria.  Using these parameter values, AMP-MMV performs either a single iteration of the parallel message schedule, or a single forward/backward pass of the serial message schedule, as described in \secref{scheduling}.  Upon completing this first iteration, approximate marginal posterior distributions are available for each of the underlying random variables, e.g., $p(\nt{x}|\ovec{y})$, $p(s_n|\ovec{y})$, and $p(\nt{\theta}|\ovec{y})$.  Additionally, belief propagation can provide pairwise joint posterior distributions, e.g., $p(\nt{\theta}, \theta_n^{(t-1)}|\ovec{y})$, for any variable nodes connected by a common factor node \cite{B2006}.  With these marginal, and pairwise joint, posterior distributions, it is possible to perform the iterative expectation and maximization steps required to maximize $p(\ovec{y} | \Gamma)$ in closed-form.  We adopt a Gauss-Seidel 
scheme, performing coordinate-wise maximization, e.g.,
\begin{equation}
	\lambda^{k+1} = \argmax_\lambda \text{E}_{\vec{s},\ovec{\theta}|\ovec{y}}\left[\log p(\ovec{y}, \vec{s}, \ovec{\theta}; \lambda, \Gamma^k \! \setminus \!\! \{\! \lambda^k \!\}) \Big| \ovec{y}, \Gamma^k \right], \nonumber
\end{equation}
where $k$ is the iteration index common to both AMP-MMV and the EM algorithm.

In \tabref{em_updates} we provide the EM parameter update equations for our signal model.  In practice, we found that the robustness and convergence behavior of our EM procedure were improved if we were selective about which parameters we updated on a given iteration.  For example, the parameters $\alpha$ and $\rho$ are tightly coupled to one another, since $\text{var}\{\nt{\theta} | \theta_n^{(t-1)}\} = \alpha^2 \rho$.  Consequently, if the initial choices of $\alpha$ and $\rho$ are too small, it is possible that the EM procedure will overcompensate on the first iteration by producing revised estimates of both parameters that are too large.  This leads to an oscillatory behavior in the EM updates that can be effectively combated by avoiding updating both $\alpha$ and $\rho$ on the same iteration.

\begin{table}
\centering
\scriptsize
\setlength{\tabcolsep}{2pt}
\setlength{\belowcaptionskip}{2ex}
\begin{tabular}{|llr|}
	\hline
	\multicolumn{3}{| l |}{$\textsf{\% Define key quantities obtained from AMP-MMV at iteration } k \textsf{:}$}\\
	$\text{E}[s_n | \ovec{y}] = \frac{\lambda_n \prod_{t=1}^T \nt{\fwd{\pi}}}{\lambda_n \prod_{t=1}^T \nt{\fwd{\pi}} + (1 - \lambda_n) \prod_{t=1}^T (1 - \nt{\fwd{\pi}})}$ & & (Q1)\\[2ex]
	$\tilde{v}_n^{(t)} \triangleq \text{var}\{\theta_n^{(t)} | \ovec{y}\} = \left(\frac{1}{\fwd{\kappa}_n^{(t)}} + \frac{1}{\fwd{\psi}_n^{(t)}} + \frac{1}{\bwd{\kappa}_n^{(t)}} \right)^{-1}$ & & (Q2)\\[2ex]
	$\tilde{\mu}_n^{(t)} \triangleq \text{E}[\theta_n^{(t)} | \ovec{y}] = \tilde{v}_n^{(t)} \cdot  \left(\frac{\fwd{\eta}_n^{(t)}}{\fwd{\kappa}_n^{(t)}} + \frac{\fwd{\xi}_n^{(t)}}{\fwd{\psi}_n^{(t)}} + \frac{\bwd{\eta}_n^{(t)}}{\bwd{\kappa}_n^{(t)}} \right)$ & & (Q3)\\[2ex]
 	$v_n^{(t)} \triangleq \text{var}\big\{x_n^{(t)} \big| \ovec{y}\big\} \qquad \textsf{\% See \eqref{amp_v_defn} of \tabref{algorithm_equations}}$ & & \\
	$\mu_n^{(t)} \triangleq \text{E}\big[x_n^{(t)} \big| \ovec{y}\big] \qquad \quad \textsf{\% See \eqref{amp_mu_defn} of \tabref{algorithm_equations}}$ & & \\[1ex]
 	\hline
	\multicolumn{3}{| l |}{$\textsf{\% EM update equations:}$}\\
	$\lambda^{k+1} = \tfrac{1}{N} \sum_{n=1}^{N} \text{E}[s_n | \ovec{y}]$ & & \emcnt{} \\[0ex]
	
	$\zeta^{k+1} = \left( \tfrac{N(T-1)}{\rho^k} + \tfrac{N}{(\sigma^2)^k} \right)^{-1} \Big( \tfrac{1}{(\sigma^2)^k} \sum_{n=1}^N \tilde{\mu}_n^{(1)}$ & & \\[1ex]
	$\qquad \quad + \sum_{t=2}^T \sum_{n=1}^N \tfrac{1}{\alpha^k \rho^k} \big(\nt{\tilde{\mu}} - (1 - \alpha^k) \tilde{\mu}_n^{(t-1)}\big) \Big)$ & & \emcnt{} \\[1ex]
	
	$\alpha^{k+1} = \tfrac{1}{4N(T-1)} \Big(\mf{b} - \sqrt{\mf{b}^2 + 8N(T-1) \mf{c}}\Big)$ & & \emcnt{} \\[1ex]
	\quad $\textsf{where:}$ & & \\[0ex]
	$\quad \mf{b} \triangleq \tfrac{2}{\rho^k} \sum_{t=2}^T \sum_{n=1}^N \mf{Re}\big\{ \text{E}[{\nt{\theta}}^{*} \theta_n^{(t-1)} | \ovec{y}] \big\}$  & & \\[0ex]
	$\qquad \qquad - \mf{Re}\{(\nt{\tilde{\mu}} - \tilde{\mu}_n^{(t-1)})^{*} \zeta^k\} - \tilde{v}_n^{(t-1)} - |\tilde{\mu}_n^{(t-1)}|^2$ & & \\[0ex]
	$\quad \mf{c} \triangleq \tfrac{2}{\rho^k} \sum_{t=2}^T \sum_{n=1}^N \nt{\tilde{v}} + |\nt{\tilde{\mu}}|^2 + \tilde{v}_n^{(t-1)} + |\tilde{\mu}_n^{(t-1)}|^2$ & & \\[0ex]
	$\qquad \qquad - 2 \mf{Re}\big\{ \text{E}[{\nt{\theta}}^{*} \theta_n^{(t-1)} | \ovec{y}] \big\}$ & & \\[0ex]
	$\rho^{k+1} = \tfrac{1}{(\alpha^k)^2 N (T-1)} \sum_{t=2}^T \sum_{n=1}^N \nt{\tilde{v}} + |\nt{\tilde{\mu}}|^2$ & & \\[0ex]
	$\qquad \qquad  + (\alpha^k)^2 |\zeta^k|^2 - 2 (1 - \alpha^k) \mf{Re}\big\{ \text{E}[{\nt{\theta}}^{*} \theta_n^{(t-1)} | \ovec{y}] \big\}$ & & \\[0ex]
	$\qquad \qquad - 2 \alpha^k \mf{Re}\big\{ \tilde{\mu}_n^{(t)*} \zeta^k \big\} + 2 \alpha^k (1-\alpha^k) \mf{Re}\big\{ \tilde{\mu}_n^{(t-1)*} \zeta^k \big\}$ & & \\[0ex]
	$\qquad \qquad + (1 - \alpha^k) (\tilde{v}_n^{(t-1)} + |\tilde{\mu}_n^{(t-1)}|^2)$ & & \emcnt{} \\[0ex]
	$\sigma_e^{2\,\,k+1} = \tfrac{1}{TM} \left( \sum_{t=1}^T \|\vect{y} - \vec{A} \vect{\mu}\|^2 + \vec{1}_{_N}^T \vect{v} \right)$ & & \emcnt{} \\[0ex]
	\hline
\end{tabular}

\caption{EM algorithm update equations for the signal model parameters of \secref{signal_model}.}
\label{tab:em_updates}
\end{table}

\section{Numerical Study}
\label{sec:numerical_study}

In this section we describe the results of an extensive numerical study that was conducted to explore the performance characteristics and tradeoffs of AMP-MMV.  MATLAB code\footnote{Code available at \url{ece.osu.edu/~schniter/turboAMPmmv}.} was written to implement both the parallel and serial message schedules of \secref{scheduling}, along with the EM parameter estimation procedure of \secref{parameter_estimation}.

For comparison to AMP-MMV, we tested two other Bayesian algorithms for the MMV problem, MSBL \cite{WR2007} and T-MSBL\footnote{Code available at \url{dsp.ucsd.edu/~zhilin/Software.html}.} \cite{ZR2011}, which have been shown to offer ``best in class'' performance on the MMV problem.  We also included a recently proposed greedy algorithm designed specifically for highly correlated signals, subspace-augmented MUSIC\footnote{Code obtained through personal correspondence with authors.} (SA-MUSIC), which has been shown to outperform MMV basis pursuit and several correlation-agnostic greedy methods \cite{LBJ2010}.  Finally, we implemented the support-aware Kalman smoother (SKS), which, as noted in \secref{kalman}, provides a lower bound on the achievable MSE of any algorithm.  To implement the SKS, we took advantage of the fact that $\ovec{y}$, $\ovec{x}$, and $\ovec{\theta}$ are jointly Gaussian when conditioned on the support, $\vec{s}$, and thus \figref{total_factor_graph} becomes a Gaussian graphical model.  Consequently, the sum-product algorithm yields closed-form expressions (i.e., no approximations are required) for each of the messages traversing the graph.  Therefore, it is possible to obtain the desired posterior means (i.e., MMSE estimates of $\ovec{x}$) despite the fact that the graph is loopy \cite[Claim 5]{WF2001}.

{In all of our experiments, performance was analyzed on synthetically generated datasets, and averaged over $250$ independent trials.  Since MSBL and T-MSBL were derived for real-valued signals, we used a real-valued equivalent of the signal model described in \secref{signal_model}, and ran a real-valued version of AMP-MMV.  Our data generation procedure closely mirrors the one used to characterize T-MSBL in \cite{ZR2011}.  Unless otherwise stated, the measurement matrices were i.i.d. Gaussian random matrices with unit-norm columns, $T = 4$ measurement vectors were generated, the stationary variance of the amplitude process was set at $\sigma^2 \triangleq \tfrac{\alpha \rho}{2 - \alpha} = 1$, and the noise variance $\sigma_e^2$ was set to yield an SNR of $25$ dB.}

Three performance metrics were considered throughout our tests.  The first metric, which we refer to as the time-averaged normalized MSE (TNMSE), is defined as
\begin{equation}
	\text{TNMSE}(\ovec{x}, \hat{\ovec{x}}) \triangleq \frac{1}{T} \sum_{t=1}^T \frac{\|\vect{x} - \hvec{x}^{(t)}\|_2^2}{\|\vect{x}\|_2^2}, \nonumber
\end{equation}
where $\hvec{x}^{(t)}$ is an estimate of $\vect{x}$.  The second metric, intended to gauge the accuracy of the recovered support, is the normalized support error rate (NSER), {which is defined as the number of indices in which the true and estimated support differ, normalized by the cardinality of the true support $\mathcal{S}$.}  The third and final metric is runtime, which is an important metric given the prevalence of high-dimensional datasets.

{The algorithms were configured and executed as follows: to obtain support estimates for MSBL, T-MSBL, and AMP-MMV, we adopted the technique utilized in \cite{ZR2011} of identifying the $K$ amplitude trajectories with the largest $\ell_2$ norms as the support set, where $K \triangleq |\mathcal{S}|$.  Note that this is an optimistic means of identifying the support, as it assumes that an oracle provides the true value of $K$.  For this reason, we implemented an additional \emph{non}-oracle-aided support estimate for AMP-MMV that consisted of those indices $n$ for which $\hat{p}(s_n|\ovec{y}) > \tfrac{1}{2}$.}  In all simulations, AMP-MMV was given imperfect knowledge of the signal model parameters, and refined the initial parameter choices according to the EM update procedure given in \tabref{em_updates}.  In particular, the noise variance was initialized at $\sigma_e^2 = 1 \times 10^{-3}$.  The remaining parameters were initialized {agnostically} using simple heuristics that made use of sample statistics derived from the available measurements, $\ovec{y}$.  {\Eqref{amp_x_hat} of \tabref{algorithm_equations} was used to produce $\hvec{x}^{(t)}$, which corresponds to an MMSE estimate of $\vect{x}$ under AMP-MMV's estimated posteriors $\hat{p}(\nt{x}|\ovec{y})$.}  In the course of running simulations, we monitored the residual energy, $\sum_{t=1}^T \|\vect{y} - \vec{A}\hvec{x}^{(t)}\|_2^2$, and would automatically switch the schedule, e.g., from parallel to serial, and/or change the maximum number of iterations whenever the residual energy exceeded a noise variance-dependent threshold.  The SKS was given perfect parameter and support knowledge {and was run until convergence}.  Both MSBL and T-MSBL were {tuned in a manner} recommended by the codes' authors.  SA-MUSIC was given the true value of $K$, and upon generating an estimate of the support, $\hat{\mathcal{S}}$, a conditional MMSE signal estimate was produced, e.g., $\hvec{x}^{(t)} = \text{E}[\vect{x} | \hat{\mathcal{S}}, \vect{y}]$.

\subsection{Performance Versus Sparsity, $M/K$}
\label{sec:numerical_study:sparsity}

As a first experiment, we studied how performance changes as a function of the measurements-to-active-coefficients ratio, $M/K$.  For this experiment, {$N = 5000$, $M = 1563$, and $T = 4$.  The activity probability, $\lambda$, was swept over the range $[0.096, 0.22]$, implying that the ratio of measurements-to-active-coefficients, $M/K$, ranged from $1.42$ to $3.26$.}

\begin{figure*}
	\begin{center}
		\scalebox{0.45}{\includegraphics*[-2.50in,3.5in][11.5in,7.75in]{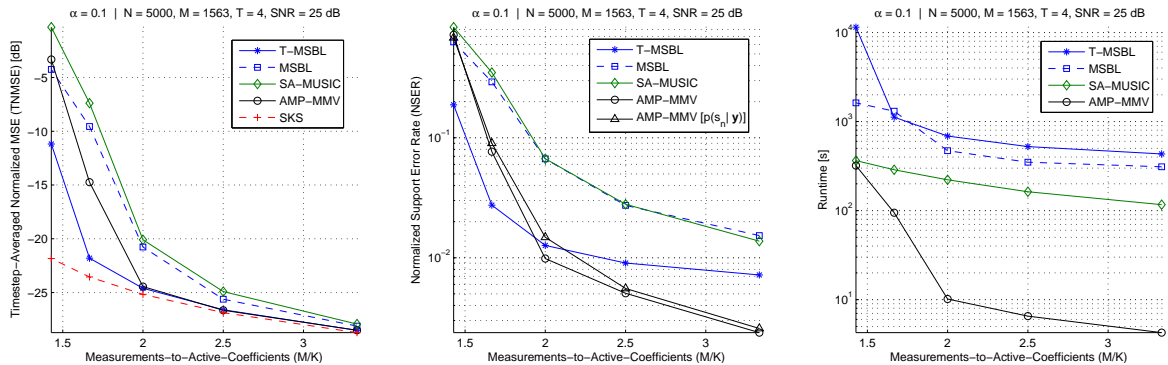}}
		\caption{A plot of the TNMSE (in dB), NSER, and runtime of T-MSBL, MSBL, SA-MUSIC, AMP-MMV, and the SKS versus $M$/$K$.  Correlation coefficient $1 - \alpha = 0.90$.}
		\label{fig:sweep_lam_alpha_0_10}
	\end{center}
\end{figure*}

In \figref{sweep_lam_alpha_0_10}, we plot the performance when the temporal correlation of the amplitudes is $1 - \alpha = 0.90$.  For AMP-MMV, two traces appear on the NSER plot, with the \small$\bigcirc$ \normalsize marker corresponding to the $K$-largest-trajectory-norm method of support estimation, and the $\bigtriangleup$ marker corresponding to the support estimate obtained from the posteriors $\hat{p}(s_n | \ovec{y})$.  We see that, when $M/K \ge 2$, the TNMSE performance of both AMP-MMV and T-MSBL is almost identical to that of the oracle-aided SKS.  However, when $M/K < 2$, every algorithm's support estimation performance (NSER) degrades, and the TNMSE consequently grows.  Indeed, when $M/K < 1.50$, all of the algorithms perform poorly compared to the SKS, although T-MSBL performs the best of the four.  We also note the superior NSER performance of AMP-MMV over much of the range, even when using $p(s_n | \ovec{y})$ to estimate $\mathcal{S}$ (and thus not requiring apriori knowledge of $K$).  From the runtime plot we see the tremendous efficiency of AMP-MMV.  Over the region in which AMP-MMV is performing well (and thus not cycling through multiple configurations in vain), we see that AMP-MMV's runtime is more than one order-of-magnitude faster than SA-MUSIC, and two orders-of-magnitude faster than either T-MSBL or MSBL.

\begin{figure*}
	\begin{center}
		\scalebox{0.45}{\includegraphics*[-2.50in,3.5in][11.5in,7.75in]{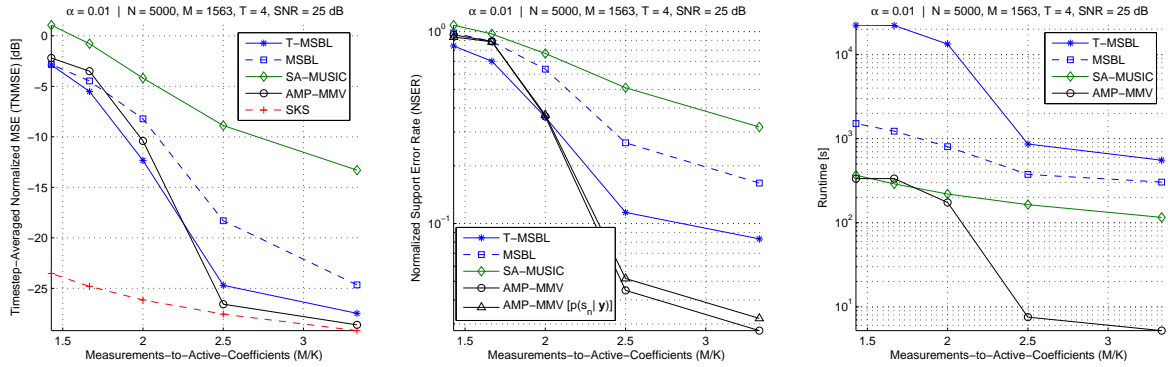}}
		\caption{A plot of the TNMSE (in dB), NSER, and runtime of T-MSBL, MSBL, SA-MUSIC, AMP-MMV, and the SKS versus $M/K$.  Correlation coefficient $1 - \alpha = 0.99$.}
		\label{fig:sweep_lam_alpha_0_01}
	\end{center}
\end{figure*}

In \figref{sweep_lam_alpha_0_01} we repeat the same experiment, but with increased amplitude correlation $1 - \alpha = 0.99$.  In this case we see that AMP-MMV and T-MSBL still offer a TNMSE performance that is comparable to the SKS when $M/K \ge 2.50$, whereas the performance of both MSBL and SA-MUSIC has degraded across-the-board.  When $M/K < 2.5$, the NSER and TNMSE performance of AMP-MMV and T-MSBL decay sharply, and all the methods considered perform poorly compared to the SKS.  Our finding that performance is adversely affected by increased temporal correlation is consistent with the theoretical and empirical findings of \cite{WR2007,ER2010,LBJ2010,ZR2011}.  Interestingly, the performance of the SKS shows a modest improvement compared to \figref{sweep_lam_alpha_0_10}, reflecting the fact that the slower temporal variations of the amplitudes are easier to track when the support is known.

\subsection{Performance Versus $T$}
\label{sec:numerical_study:timestep}

\begin{figure*}
	\begin{center}
		\scalebox{0.45}{\includegraphics*[-2.50in,3.5in][11.5in,7.75in]{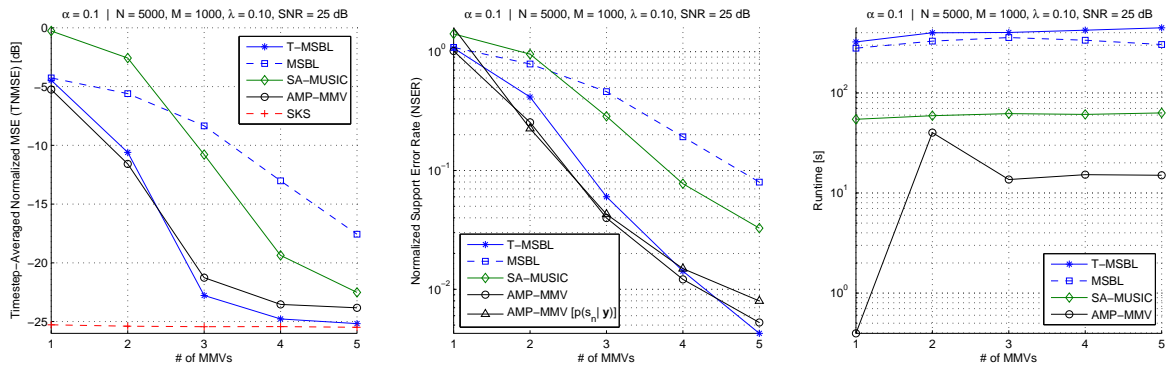}}
		\caption{A plot of the TNMSE (in dB), NSER, and runtime of T-MSBL, MSBL, SA-MUSIC, AMP-MMV, and the SKS versus $T$.  Correlation coefficient 1 - $\alpha = 0.90$.}
		\label{fig:sweep_T_alpha_0_10}
	\end{center}
\end{figure*}

In a second experiment, we studied how performance is affected by the number of measurement vectors, $T$, used in the reconstruction.  For this experiment, we used $N = 5000$, $M = N / 5$, and $\lambda = 0.10$ ($M/K = 2$).  \Figref{sweep_T_alpha_0_10} shows the performance with a correlation of $1 - \alpha = 0.90$.  Comparing to \figref{sweep_lam_alpha_0_10}, we see that MSBL's performance is strongly impacted by the reduced value of $M$.  AMP-MMV and T-MSBL perform more-or-less equivalently across the range of $T$, although AMP-MMV does so with an order-of-magnitude reduction in complexity.  It is interesting to observe that, in this problem regime, the SKS TNMSE bound is insensitive to the number of measurement vectors acquired.

\subsection{Performance Versus SNR}
\label{sec:numerical_study:snr}

To understand how AMP-MMV performs in low SNR environments, we conducted a test in which SNR was swept from $5$ dB to $25$ dB.\footnote{In lower SNR regimes, learning rules for the noise variance are known to become less reliable \cite{WR2007,ZR2011}.  Still, for high-dimensional problems, a sub-optimal learning rule may be preferable to a computationally costly cross-validation procedure.  For this reason, we ran all three Bayesian algorithms with a learning rule for the noise variance enabled.}  The problem dimensions were fixed at $N = 5000$, $M = N/5$, and $T = 4$.  The sparsity rate, $\lambda$, was chosen to yield $M/K = 3$ measurements-per-active-coefficient, and the correlation was set at $1 - \alpha = 0.95$.

\begin{figure*}
	\begin{center}
		\scalebox{0.45}{\includegraphics*[-2.50in,3.5in][11.5in,7.75in]{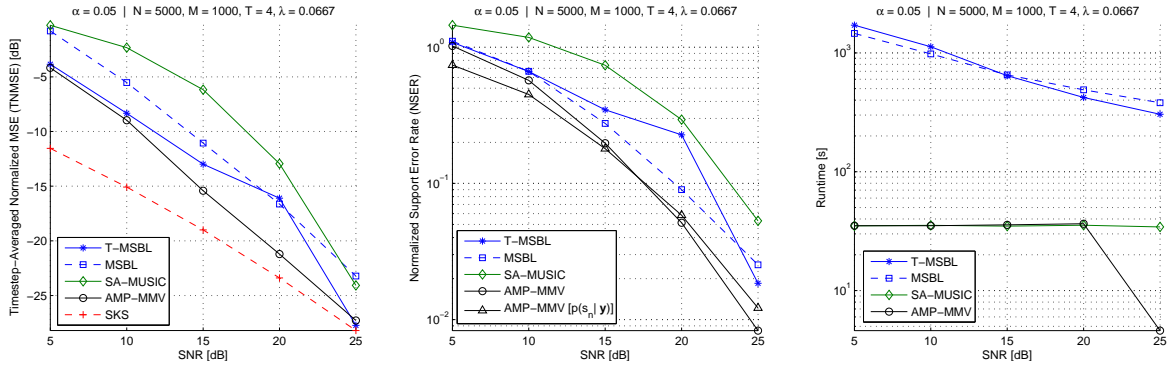}}
		\caption{A plot of the TNMSE (in dB), NSER, and runtime of T-MSBL, MSBL, SA-MUSIC, AMP-MMV, and the SKS versus SNR.  Correlation coefficient $1 - \alpha = 0.95$.}
		\label{fig:sweep_SNR_alpha_0_05}
	\end{center}
\end{figure*}

Our findings are presented in \figref{sweep_SNR_alpha_0_05}.  Both T-MSBL and MSBL operate within $5$ - $10$ dB of the SKS in TNMSE across the range of SNRs, while AMP-MMV operates $\approx 5$ dB from the SKS when the SNR is at or below $10$ dB, and approaches the SKS in performance as the SNR elevates.  We also note that using AMP-MMV's posteriors on $s_n$ to estimate the support does not appear to perform much worse than the $K$-largest-trajectory-norm method for high SNRs, and shows a slight advantage at low SNRs.  The increase in runtime exhibited by AMP-MMV in this experiment is a consequence of our decision to configure AMP-MMV identically for all experiments; our initialization of the noise variance, $\sigma_e^2$, was more than an order-of-magnitude off over the majority of the SNR range, and thus AMP-MMV cycled through many different schedules in an effort to obtain an (unrealistic) residual energy.  Runtime could be drastically improved in this experiment by using a more appropriate initialization of $\sigma_e^2$.

\subsection{Performance Versus Undersampling Rate, $N/M$}
\label{sec:numerical_study:undersampling}

\begin{figure*}
	\begin{center}
		\scalebox{0.45}{\includegraphics*[-2.50in,3.5in][11.5in,7.75in]{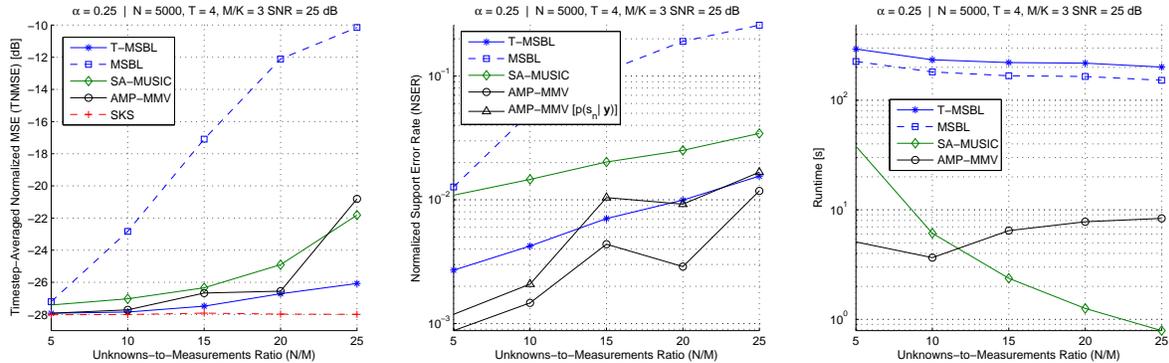}}
		\caption{A plot of the TNMSE (in dB), NSER, and runtime of T-MSBL, MSBL, SA-MUSIC, AMP-MMV, and the SKS versus undersampling rate, $N/M$.  Correlation coefficient $1 - \alpha = 0.75$.}
		\label{fig:sweep_undersampling_alpha_0_25}
	\end{center}
\end{figure*}

As mentioned in \secref{introduction}, one of the principal aims of CS is to reduce the number of measurements that must be acquired while still obtaining a good solution.  In the MMV problem, dramatic reductions in the sampling rate are possible.  To illustrate this, in \figref{sweep_undersampling_alpha_0_25} we present the results of an experiment in which the undersampling factor, $N/M$, was varied from $5$ to $25$ unknowns-per-measurement.  Specifically, $N$ was fixed at $5000$, while $M$ was varied.  $\lambda$ was likewise adjusted in order to keep $M$/$K$ fixed at $3$ measurements-per-active-coefficient.  In \figref{sweep_undersampling_alpha_0_25}, we see that MSBL quickly departs from the SKS performance bound, whereas AMP-MMV, T-MSBL, and SA-MUSIC are able to remain close to the bound when $N/M \le 20$.  At $N/M = 25$, both AMP-MMV and SA-MUSIC have diverged from the bound, and, while still offering an impressive TNMSE, they are outperformed by T-MSBL.  In conducting this test, we observed that AMP-MMV's performance is strongly tied to the number of smoothing iterations performed.  Whereas for other tests, $5$ smoothing iterations were often sufficient, in scenarios with a high degree of undersampling, (e.g., $N/M \ge 15$), $50-100$ smoothing iterations were often required to obtain good signal estimates.  This suggests that messages must be exchanged between neighboring timesteps over many iterations in order to arrive at consensus in severely underdetermined problems.

\subsection{Performance Versus Signal Dimension, $N$}
\label{sec:numerical_study:dimension}

As we have indicated throughout this paper, a key consideration of our method was ensuring that it would be suitable for high-dimensional problems.  Our complexity analysis indicated that a single iteration of AMP-MMV could be completed in $\mathcal{O}(TNM)$ flops.  This linear scaling of the complexity with respect to problem dimensions gives encouragement that our algorithm should efficiently handle large problems, but if the number of iterations required to obtain a solution grows too rapidly with problem size, our technique would be of limited practical utility.  To ensure that this was not the case, we performed an experiment in which the signal dimension, $N$, was swept logarithmically over the range $[100, 10000]$.  $M$ was scaled proportionally such that $N/M = 3$.  The sparsity rate was fixed at $\lambda = 0.15$ so that $M/K \approx 2$, and the correlation was set at $1 - \alpha = 0.95$.

\begin{figure*}
	\begin{center}
		\scalebox{0.45}{\includegraphics*[-2.50in,3.45in][11.5in,7.75in]{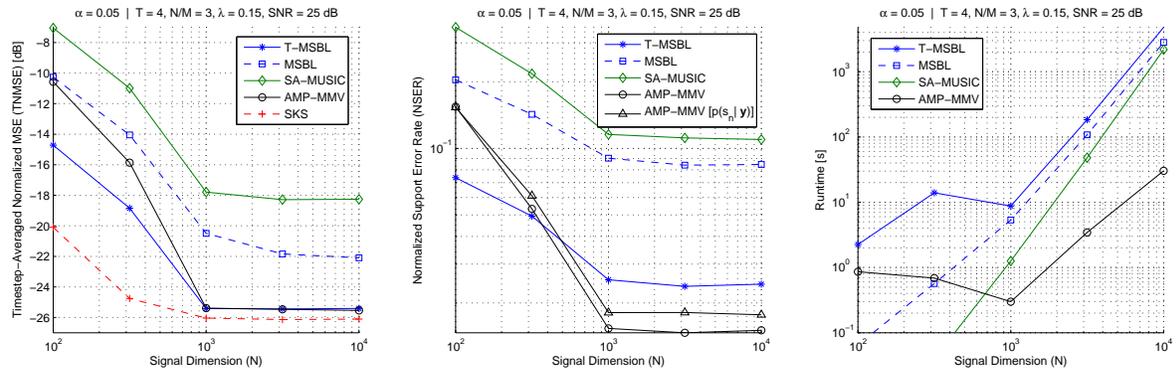}}
		\caption{A plot of the TNMSE (in dB), NSER, and runtime of T-MSBL, MSBL, SA-MUSIC, AMP-MMV, and the SKS versus signal dimension, $N$.  Correlation coefficient $1 - \alpha = 0.95$.}
		\label{fig:sweep_dimension_alpha_0_05}
	\end{center}
\end{figure*}

The results of this experiment are provided in \figref{sweep_dimension_alpha_0_05}. Several features of these plots are of interest.  First, we observe that the performance of every algorithm improves noticeably as problem dimensions grow from $N = 100$ to $N = 1000$, with AMP-MMV and T-MSBL converging in TNMSE performance to the SKS bound.  The second observation that we point out is that AMP-MMV works extremely quickly.  Indeed, a problem with $NT = 40000$ unknowns can be solved accurately in just under $30$ seconds.  Finally, we note that at small problem dimensions, AMP-MMV is not as quick as either MSBL or SA-MUSIC, however AMP-MMV scales with increasing problem dimensions more favorably than the other methods; at $N = 10000$ we note that AMP-MMV runs at least two orders-of-magnitude faster than the other techniques.

\subsection{Performance With Time-Varying Measurement Matrices}
\label{sec:numerical_study:matrices}

In all of the previous experiments, we considered the standard MMV problem \eqref{linear_model}, in which all of the measurement vectors were acquired using a single, common measurement matrix.  While this setup is appropriate for many tasks, there are a number of practical applications in which a joint-sparse signal is measured through distinct measurement matrices.

To better understand what, if any, gains can be obtained from diversity in the measurement matrices, we designed an experiment that explored how performance is affected by the rate-of-change of the measurement matrix over time.  For simplicity, we considered a first-order Gauss-Markov random process to describe how a given measurement matrix changed over time.  Specifically, we started with a matrix whose columns were drawn i.i.d. Gaussian as in previous experiements, which was then used as the measurement matrix to collect the measurements at timestep $t=1$.  At subsequent timesteps, the matrix evolved according to
\begin{equation}
	\vect{A} = (1 - \beta) \vec{A}^{(t-1)} + \beta \vect{U},
	\label{eq:mtx_evolve}
\end{equation}
where $\vect{U}$ was a matrix whose elements were drawn i.i.d. Gaussian, with a variance chosen such that the column norm of $\vect{A}$ would (in expectation) equal one.

In the test, $\beta$ was swept over a range, providing a quantitative measure of the rate-of-change of the measurement matrix over time.  Clearly, $\beta = 0$ would correspond to the standard MMV problem, while $\beta = 1$ would represent a collection of statistically independent measurement matrices.

\begin{figure*}
	\begin{center}
		\scalebox{0.45}{\includegraphics*[-2.50in,3.4in][11.5in,7.75in]{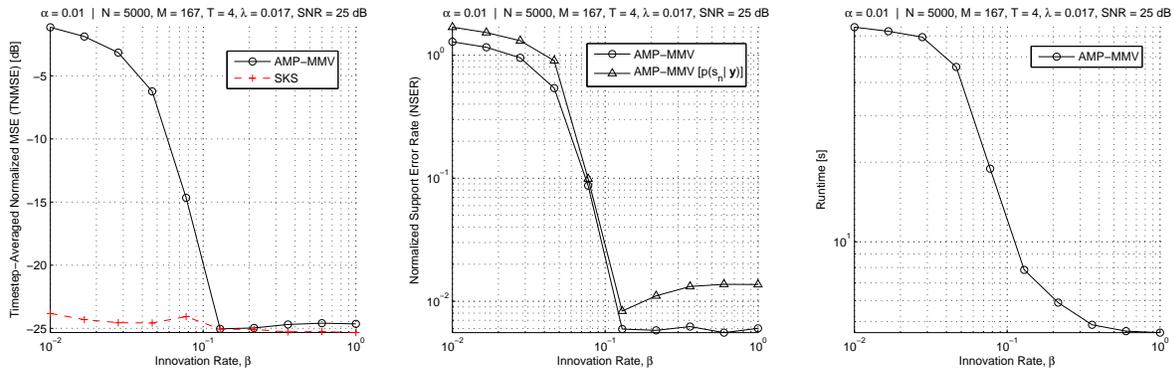}}
		\caption{A plot of the TNMSE (in dB), NSER, and runtime of AMP-MMV and the SKS versus rate-of-change of the measurement matrix, $\beta$.  Correlation coefficient $1 - \alpha = 0.99$.}
		\label{fig:change_mtx_alpha_0_01}
	\end{center}
\end{figure*}

In \figref{change_mtx_alpha_0_01} we show the performance when $N = 5000$, $N/M = 30$, $M/K = 2$, and the correlation is $1 - \alpha = 0.99$.  For the standard MMV problem, this configuration is effectively impossible.  Indeed, for $\beta < 0.03$, we see that AMP-MMV is entirely failing at recovering the signal.  However, once $\beta \approx 0.08$, we see that the NSER has dropped dramatically, as has the TNMSE.  Once $\beta \ge 0.10$, AMP-MMV is performing almost to the level of the noise.  As this experiment should hopefully convince the reader, even modest amounts of diversity in the measurement process can enable accurate reconstruction in operating environments that are otherwise impossible.

\section{Conclusion}
\label{sec:conclusion}
In this work we introduced AMP-MMV, a Bayesian message passing algorithm for solving the MMV problem \eqref{linear_model} when temporal correlation is present in the amplitudes of the non-zero signal coefficients.  Our algorithm, which leverages Donoho, Maleki, and Montanari's AMP framework \cite{DMM2009}, performs rapid inference on high-dimensional MMV datasets.  In order to establish a reference point for the quality of solutions obtained by AMP-MMV, we described and implemented the oracle-aided support-aware Kalman smoother (SKS).  In numerical experiments, we found a range of problems over which AMP-MMV performed nearly as well as the SKS, despite the fact that AMP-MMV was given crude hyperparameter initializations that were refined from the data using an expectation-maximization algorithm.  In comparing against two alternative Bayesian techniques, and one greedy technique, we found that AMP-MMV offers {an unrivaled performance-complexity tradeoff}, particular in high-dimensional settings.  We also demonstrated that substantial gains can be obtained in the MMV problem by incorporating diversity into the measurement process.  Such diversity is particularly important in settings where the temporal correlation between coefficient amplitudes is substantial.

\appendices

\section{Taylor Series Approximation of $\msg{\nt{f}}{\nt{\theta}}^{\text{mod}}$}
\label{app:taylor}
In this appendix we summarize the procedure used to collapse the binary Gaussian mixture of \eqref{f_to_theta_msg_mod}, $\msg{\nt{f}}{\nt{\theta}}^{\text{mod}}(\nt{\theta})$, to a single Gaussian, $\msg{\nt{f}}{\nt{\theta}}(\nt{\theta}) = \mathcal{CN}(\nt{\theta}; \nt{\fwd{\xi}}, \nt{\fwd{\psi}})$.  For simplicity, we drop the $n$ and $(t)$ sub- and superscripts.

Let $\theta_r \triangleq \mathfrak{Re}\{\theta\}$, let $\theta_i \triangleq \mathfrak{Im}\{\theta\}$, and let $\phi_r$ and $\phi_i$ be defined similarly.  Define
\ifthenelse{\boolean{ONE_COLUMN}}
{
\begin{eqnarray}
	\tilde{g}(\theta_r, \theta_i) &\triangleq& \msg{f}{\theta}^{\text{mod}}(\theta_r + j \theta_i),	\nonumber \\
	&=& (1 - \Omega(\bwd{\pi})) \,\, \mathcal{CN}(\theta_r + j \theta_i; \tfrac{1}{\eps} \phi, \tfrac{1}{\eps^2} c) + \Omega(\bwd{\pi}) \,\, \mathcal{CN}(\theta_r + j \theta_i; \phi, c)	\nonumber \\
	\tilde{f}(\theta_r, \theta_i) &\triangleq& - \log \tilde{g}(\theta_r, \theta_i)	\nonumber.
\end{eqnarray}
}
{
\begin{eqnarray}
	\tilde{g}(\theta_r, \theta_i) &\triangleq& \msg{f}{\theta}^{\text{mod}}(\theta_r + j \theta_i),	\nonumber \\
	&=& (1 - \Omega(\bwd{\pi})) \,\, \mathcal{CN}(\theta_r + j \theta_i; \tfrac{1}{\eps} \phi, \tfrac{1}{\eps^2} c)	\nonumber \\
	&\quad& + \,\, \Omega(\bwd{\pi}) \,\, \mathcal{CN}(\theta_r + j \theta_i; \phi, c)	\nonumber \\
	\tilde{f}(\theta_r, \theta_i) &\triangleq& - \log \tilde{g}(\theta_r, \theta_i)	\nonumber.
\end{eqnarray}
}
Our objective is to approximate $\tilde{f}(\theta_r, \theta_i)$ using a two-dimensional second-order Taylor series expansion, $\breve{f}(\theta_r, \theta_i)$, about the point $\phi$:
\ifthenelse{\boolean{ONE_COLUMN}}
{
\begin{eqnarray}
	\breve{f}(\theta_r, \theta_i) &=& \tilde{f}(\phi_r, \phi_i) + (\theta_r - \phi_r)\left.\frac{\partial \tilde{f}}{\partial \theta_r}\right|_{\theta = \phi} + (\theta_i- \phi_i)\left.\frac{\partial \tilde{f}}{\partial \theta_i}\right|_{\theta = \phi}	\nonumber \\
	&\quad& + \frac{1}{2} \left[ (\theta_r - \phi_r)^2\left.\frac{\partial^2 \tilde{f}}{\partial \theta_r^2}\right|_{\theta = \phi} + (\theta_r - \phi_r)(\theta_i - \phi_i)\left.\frac{\partial^2 \tilde{f}}{\partial \theta_r \partial \theta_i}\right|_{\theta = \phi} + (\theta_i- \phi_i)^2\left.\frac{\partial^2 \tilde{f}}{\partial \theta_i^2}\right|_{\theta = \phi} \right] .	\nonumber
\end{eqnarray}
}
{
\begin{eqnarray}
	\breve{f}(\theta_r, \theta_i) &=& \tilde{f}(\phi_r, \phi_i) + (\theta_r - \phi_r)\left.\frac{\partial \tilde{f}}{\partial \theta_r}\right|_{\theta = \phi} \nonumber \\
	&\quad& + (\theta_i- \phi_i)\left.\frac{\partial \tilde{f}}{\partial \theta_i}\right|_{\theta = \phi}	 + \frac{1}{2} \left[ (\theta_r - \phi_r)^2\left.\frac{\partial^2 \tilde{f}}{\partial \theta_r^2}\right|_{\theta = \phi} \right.	\nonumber \\
	&\quad& + (\theta_r - \phi_r)(\theta_i - \phi_i)\left.\frac{\partial^2 \tilde{f}}{\partial \theta_r \partial \theta_i}\right|_{\theta = \phi} \nonumber \\
	&\quad& \left. + (\theta_i- \phi_i)^2\left.\frac{\partial^2 \tilde{f}}{\partial \theta_i^2}\right|_{\theta = \phi} \right] .	\nonumber
\end{eqnarray}
}
It can be shown that, for Taylor series expansions about the point $\phi$, $\tfrac{\partial^2 f}{\partial \theta_r \partial \theta_i} = \mathcal{O}(\eps^2)$ and $\left|\frac{\partial^2 f}{\partial \theta_r^2} - \frac{\partial^2 f}{\partial \theta_i^2}\right| = \mathcal{O}(\eps^2)$.  Since $\eps \ll 1$, it is reasonable to therefore adopt a further approximation and assume $\tfrac{\partial^2 \tilde{f}}{\partial \theta_r \partial \theta_i} = 0$ and $\frac{\partial^2 \tilde{f}}{\partial \theta_r^2} = \frac{\partial^2 \tilde{f}}{\partial \theta_i^2}$.  With this approximation, note that
\begin{equation}
	\exp(- \breve{f}(\theta_r, \theta_i)) \propto \mathcal{CN}(\theta_r + j \theta_i; \fwd{\xi}, \fwd{\psi}),	\nonumber
\end{equation}
with
\begin{eqnarray}
	\fwd{\psi} &\triangleq& 2 \left.\frac{\partial^2 \tilde{f}}{\partial \theta_r^2}\right|_{\theta=\phi}^{-1},	\label{eq:psi_taylor} \\
	\fwd{\xi} &\triangleq& \left( \phi_r - \frac{1}{2} \fwd{\psi} \times \left.\frac{\partial \tilde{f}}{\partial \theta_r}\right|_{\theta = \phi} \right)	+ j \left( \phi_i - \frac{1}{2} \fwd{\psi} \times \left.\frac{\partial \tilde{f}}{\partial \theta_i}\right|_{\theta = \phi} \right).	\label{eq:xi_taylor}
\end{eqnarray}
The pseudocode function, $\textsf{taylor\_approx}$, that computes \eqref{psi_taylor}, \eqref{xi_taylor} given the parameters of $\msg{f}{\theta}^{\text{mod}}(\cdot)$ is provided in \tabref{taylor_eqs}.

\bibliographystyle{ieeetr}
\bibliography{time_evolve_sparse.bib}

\end{document}